\documentclass[prc]{revtex4}
\usepackage{graphicx}

\def\bvec#1{\mbox{\boldmath $#1$}}

\newcommand{\del}[2]{\frac{\partial #1}{\partial #2}}

\newcommand{\bra}{\langle}
\newcommand{\ket}{\rangle}


\begin{document}

\title{Continuum particle-vibration coupling method 
in coordinate-space representation for finite nuclei}

\author{Kazuhito Mizuyama$^{1}$}
\author{Gianluca Col\`o$^{2,1}$}
\author{Enrico Vigezzi$^{1}$}

\email{mizukazu147@gmail.com, colo@mi.infn.it, vigezzi@mi.infn.it}

\affiliation{
$^{1}$ 
INFN, Sezione di Milano, via Celoria 16, 20133 Milano (Italy)
\\
$^{2}$ 
Dipartimento di Fisica, Universit$\grave{a}$ degli 
Studi di Milano, via Celoria 16, 20133 Milano (Italy)
}

\date{\today}

\begin{abstract}
In this paper we present a new formalism to implement 
the nuclear particle-vibration coupling (PVC) model. 
The key issue is the proper treatment of the continuum,
that is allowed by the coordinate space representation. 
Our formalism,  based on the use of zero-range interactions like
the Skyrme forces, is microscopic and fully self-consistent. 
We apply it to the case of neutron
single-particle states in $^{40}$Ca, $^{208}$Pb and 
$^{24}$O. The first two cases are meant to illustrate
the comparison with the usual (i.e., discrete) PVC
model. However, we stress that the present approach allows to 
calculate properly the effect of PVC on resonant
states. We compare our results with those from
experiments in which the particle transfer
in the continuum region has been attempted. 
The latter case, namely $^{24}$O, is chosen as
an example of a weakly-bound system. 
Such a nucleus, being double-magic and
not displaying collective low-lying vibrational excitations, 
is characterized by quite pure neutron single-particle states 
around the Fermi surface.
\end{abstract}

\maketitle

\section{Introduction}

The accurate description of the single-particle (s.p.)
strength in atomic nuclei is, to a large extent, an open 
issue (for a recent discussion see, e.g., Ref. \cite{open}).
Whereas in light nuclei either {\em ab-initio} or
shell model calculations are feasible, in the case
of medium-heavy nuclei we miss a fully microscopic
theory that is able to account for the experimental findings.
Modern self-consistent models
(either based on the mean-field Hamiltonians or on some
implementation of Density Functional Theory) do
not reproduce, as a rule, the level density around the 
Fermi surface. The reader can see, as a recent example, 
the results shown in Ref. \cite{stoitsov}.  
Moreover, the fragmentation of the s.p. strength is
by definition outside the framework of those models.

In the past decades, much emphasis has been put on
the impact on the s.p. properties provided by the
coupling with various collective nuclear motions. 
The basic ideas leading to particle-vibration 
coupling (PVC) models 
in spherical nuclei, or particle-rotation coupling
models  
in deformed systems, have been discussed in textbooks 
\cite{BMII}. These couplings provide dynamical
content to the standard shell model, in keeping
with the fact that the average potential becomes
nonlocal in time or, in other words, frequency- 
or energy-dependent. We will call self-energy, in
what follows, the dynamical part of the mean potential
arising from vibrational coupling. This contribution
will be added to the static Hartree-Fock (HF) 
potential. In this way, one may be 
able to describe the fragmentation and the related
spectroscopic factors of the s.p. states, their
density (which is proportional to the effective mass
$m^*$ near the Fermi energy), the s.p. spreading widths, 
and the imaginary component of the optical potential.

In, e.g., the review article \cite{CM} one can find
a detailed discussion about the points mentioned  
in the previous paragraph, 
together with the relevant equations and the results 
of many calculations performed in the 80's for the 
single-particle strength (mainly in $^{208}$Pb). 
These calculations are mostly not self-consistent 
and it is hard to extract from them quantitative 
conclusions because of the various approximations 
involved. Certainly, they all agree qualitatively
in pointing out that PVC plays a decisive role 
to bring the density of levels near the Fermi
energy in better agreement with experiment or, 
in other words, the effective mass $m^*$ close
to the empirical value $m^*\approx m$. 

This enhancement of the effective mass around 
the Fermi energy, as compared to the HF value, is 
only one example of a phenomenon that can be explained by
assuming that single-particle and vibrational degrees 
of freedom are not independent. Other examples, 
although not treated in the current work, are worth to be
mentioned in this Introduction. Several works have 
identified the exchange of vibrational quanta (phonons) 
between particles as one important mechanism responsible
for nuclear pairing \cite{gori,kamerdjev}. In the approach 
of \cite{pastore,idini} and references therein, one aims at 
explaining the properties of superfluid nuclei by taking 
into account both the pairing induced by the phonon
exchange and the self-energy mentioned above (cf. also
the discussion in Refs. \cite{duguet,baldo}.
We also remark that important developments
are under way aiming at implementing {\em ab-initio} 
calculation schemes for open shell nuclei, 
based on self-consistent Green's functions \cite{soma} or
on the unitary correlation operator method \cite{ucom}).
Along 
the same line, more complicated processes can be 
explained by starting from elementary single-particle and
vibrational degrees of freedom, and treating their coupling 
within the framework of an appropriate field theory: the 
spreading width of nuclear giant resonances, or the
anharmonicity of two-phonon states (to mention only 
a few examples). The development of such a general 
many-body perturbation theory scheme could 
not avoid, so far, to resort to various
approximations. In particular most of the calculations 
have employed simple, phenomenological coupling 
Hamiltonians.

Recently, in order to calculate the s.p. 
strength, microscopic PVC calculations have 
become available, either based on the 
nonrelativistic Skyrme Hamiltonian \cite{colo10} 
or on Relativistic Mean Field (RMF) 
parameterizations \cite{litvinova,anatolj}.
The results seem to be satisfactory, in a 
qualitative or semi-quantitative sense, as
they point to an increase of the effective
mass around the Fermi energy. The results are
clearly sensitive to the collectivity of the
low-lying phonons produced by the self-consistent
calculations. It is still unclear whether the
results will eventually be improved by a re-fitting
of the effective interactions or by the inclusion
of higher-order processes. 
 
One of the limitations of all the PVC models
that have been introduced so far, 
lies in the fact that they discretize the s.p.
continuum (clearly, this means that the
description of the vibrations themselves relies
on the same approximation). Although in Ref. 
\cite{varenna} a scheme to calculate the self-energy 
in coordinate-space representation had been 
proposed, there is at present no available result for the 
s.p. strength (let alone more complex physical observables)
that avoids the continuum discretization.
Consequently, the goal of the present work is to introduce 
for the first time a consistent description of PVC 
with a proper treatment of the continuum. 
In order to achieve this, 
the coordinate space representation is used. 
The current work is based on previous experience
on how to treat the continuum within the
linear response theory or Random Phase 
Approximation (RPA) (see 
Refs. \cite{matsuo1,mizuyama,Bertsch1,Bertsch2,Liu}).

The outline of the paper is as follows. In Sec. 
\ref{formalism}, we describe our formalism starting
from the general formulation and stressing the
implementation of proper continuum treatment. The
main goal of the section is to display the equations
that we have implemented and solved, and discuss
the s.p. level density. In Sec. \ref{results}, the
results for our three nuclei of choice are presented
and discussed; whenever possibile, they are compared
with experimental data. Finally, we summarize the
paper and draw our conclusions in Sec. \ref{summary}.
Some details of the calculations are shown in a
few Appendices.   

\section{Formalism}\label{formalism}

\subsection{Dyson equation in coodinate space representation.}
\label{IIIA}

The particle-vibration coupling (PVC) Hamiltonian 
\cite{CM,Fetter,Richard} in coordinate space can be written as 
\begin{eqnarray}
\hat{H}_{PVC}
=
\int \!\!\!d\bvec{r}\ 
\delta\hat{\rho}(\bvec{r})
\kappa(\bvec{r})
\sum_\sigma
\hat{\psi}^\dagger(\bvec{r}\sigma)
\hat{\psi}(\bvec{r}\sigma).
\nonumber\\
\end{eqnarray}
The density variation operator 
$\delta\hat{\rho}(\bvec{r})\equiv 
\hat{\rho}(\bvec{r}) - \langle \hat{\rho}(\bvec{r})
\rangle$ (where the brackets denote the ground-state
expectation value) in second quantized form is given by
\begin{eqnarray}
\delta\hat{\rho}(\bvec{r})
=
\sum_{n\lambda}
\left[
\delta\rho_{n\lambda}(\bvec{r})\hat{\Gamma}^\dagger_{n\lambda}
+
\delta\rho^*_{n\lambda}(\bvec{r})\hat{\Gamma}_{n\lambda}
\right],
\end{eqnarray}
where $\hat{\Gamma}^\dagger_{n\lambda}$ and $\hat{\Gamma}_{n\lambda}$ 
are the creation and annihilation operators, respectively, of a 
phonon $n$ having multipolarity $\lambda$ and 
$\delta\rho_{n\lambda}$ is the corresponding transition density, 
whereas $\kappa$ is the residual force.

If the total Hamiltonian is $\hat{H}=\hat{H}_0+\hat{H}_{PVC}$,
where the term $\hat{H}_0$ describes uncoupled
s.p. states and vibrations, the many-body perturbation
theory \cite{Fetter,Richard} can be applied. In particular, 
we assume that $\hat H_{0}$ includes the HF 
Hamiltonian for the nucleons 
and the independent boson Hamiltonian for the
phonons (based on their RPA energies). We treat the term 
$\hat{H}_{PVC}$ as a perturbation using the interaction
picture. We define Green's functions in space-time representation
and we apply standard tools like the 
Wick's theorem 
to obtain the Dyson equation in terms of 
the unperturbed HF Green's function $G_0$ and the  
perturbed Green's functions $G$:
\begin{eqnarray}
G(\bvec{r}\sigma t,\bvec{r}'\sigma't')
=
G_0(\bvec{r}\sigma t,\bvec{r}'\sigma't')
+
\sum_{\sigma_1\sigma_2}
\int\!\!\!\int \!\!\!
dt_1dt_2
\int\!\!\!\int \!\!\!
d\bvec{r}_1 d\bvec{r}_2
G_0(\bvec{r}\sigma t,\bvec{r}_1\sigma_1t_1)
\Sigma(\bvec{r}_1\sigma_1t_1,\bvec{r}_2\sigma_2t_2)
G(\bvec{r}_2\sigma_2t_2,\bvec{r}'\sigma't').
\label{hfdyson1}
\end{eqnarray}
The HF Green's function satisfies $(\omega-\hat{h}_0)G_0=1$, where  
$\hat{h}_0$ is the s.p. HF Hamiltonian.
The self-energy function is defined by
\begin{eqnarray}
\Sigma(\bvec{r}_1\sigma_1t_1,\bvec{r}_2\sigma_2t_2)
=
\kappa(\bvec{r}_1)
G(\bvec{r}_1\sigma_1 t_1,\bvec{r}_2\sigma_2t_2)
\kappa(\bvec{r}_2)
iR(\bvec{r}_1t_1,\bvec{r}_2t_2),
\label{selfenrt}
\end{eqnarray}
where $R(\bvec{r}_1t_1\bvec{r}_2t_2)$ is the RPA response function 
(or phonon propagator) in the space-time representation, and 
is defined by \cite{Fetter}
\begin{eqnarray}
iR(\bvec{r}t,\bvec{r}'t')
=\bra \Psi_{RPA}|\mbox{T}
\{\delta\hat{\rho}(\bvec{r}t)\delta\hat{\rho}(\bvec{r}'t')\}
|\Psi_{RPA}\ket,
\end{eqnarray}
where $T$ denotes the time-ordered product and the formula
stresses that the phonons are defined using the RPA 
vacuum $|\Psi_{RPA}\ket$ since this is exactly the phonon
vacuum. We also note that the use of the Wick's theorem in 
the derivation of Eq. (\ref{hfdyson1}) 
implies the use of the {\it causal} representation
of the Green's functions $G$ and $G_0$, as well as of the RPA 
response function $R$. The connection between the causal representation 
with the retarded and advanced representations is outlined in the 
Appendices, where the causal functions will
be denoted by $G^C,R^C$. This label will be omitted in the main text,
where we shall only make use of the causal functions. 

\begin{figure}
\includegraphics[width=0.6\linewidth]{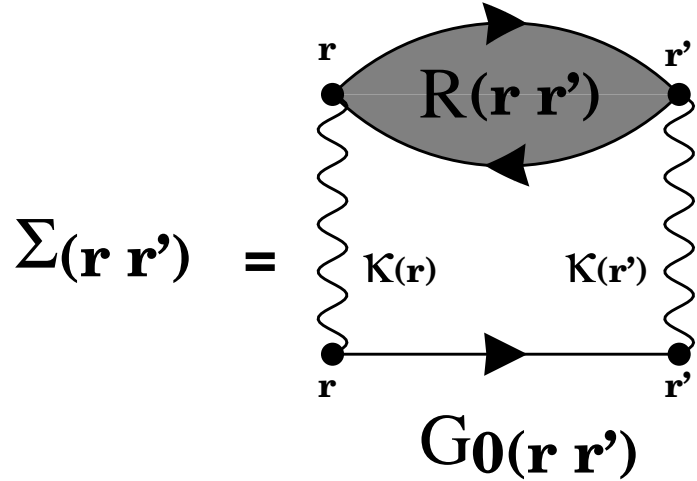}
\caption{The Feynman diagram for the self-energy function 
which corresponds to Eq. (\ref{selfen}), in the approximation
in which $G_0$ replaces $G$.}
\label{selfendiag}
\end{figure}

The Fourier transform of Eq. (\ref{hfdyson1}) is given by
\begin{eqnarray}
G(\bvec{r}\sigma,\bvec{r}'\sigma';\omega)
=
G_0(\bvec{r}\sigma,\bvec{r}'\sigma';\omega)
+
\sum_{\sigma_1\sigma_2}
\int\!\!\!\int \!\!\!d\bvec{r}_1 d\bvec{r}_2
G_0(\bvec{r}\sigma,\bvec{r}_1\sigma_1;\omega)
\Sigma(\bvec{r}_1\sigma_1,\bvec{r}_2\sigma_2;\omega)
G(\bvec{r}_2\sigma_2,\bvec{r}'\sigma';\omega), 
\label{hfdyson2}
\end{eqnarray}
while the Fourier transform of the self-energy reads
\begin{eqnarray}
&&
\Sigma(\bvec{r}\sigma,\bvec{r}'\sigma';\omega)
=
\int_{-\infty}^{\infty}\frac{d\omega'}{2\pi}
\kappa(\bvec{r})
G(\bvec{r}\sigma,\bvec{r}'\sigma';\omega-\omega')
\kappa(\bvec{r}')
iR(\bvec{r}\bvec{r}';\omega')
\label{selfen}
\end{eqnarray}
due to the convolution theorem. 

A self-consistent solution of the Dyson equation 
involves the iteration of 
the two previous equations until convergence is reached.
In practice, this is almost never done. In our work, 
since we explore for the first time the proper
continuum coupling, we limit ourselves to the
first iteration by replacing 
$G$ with $G_0$ in Eq. 
(\ref{selfen}). 

We restrict our investigation to spherical systems
in which static pairing correlations vanish. By taking 
profit of the spherical symmetry, one can use partial wave 
expansions and arrive at 
\begin{eqnarray}
\Sigma(\bvec{r}\sigma,\bvec{r}'\sigma';\omega)
&=&
\sum_{ljm}
\frac{\mathcal{Y}_{ljm}(\hat{\bvec{r}}\sigma)}{r}
\Sigma_{lj}(rr';\omega)
\frac{\mathcal{Y}^*_{ljm}(\hat{\bvec{r}}'\sigma')}{r'},
\nonumber\\
\\
G(\bvec{r}\sigma,\bvec{r}'\sigma';\omega)
&=&
\sum_{ljm}
\frac{\mathcal{Y}_{ljm}(\hat{\bvec{r}}\sigma)}{r}
G_{lj}(rr';\omega)
\frac{\mathcal{Y}^*_{ljm}(\hat{\bvec{r}}'\sigma')}{r'},
\nonumber\\
\end{eqnarray}
where $\mathcal{Y}_{ljm}(\hat{\bvec{r}}\sigma) \equiv
\left[ Y_{l}(\hat{\bvec{r}}) \otimes \chi_{1/2}(\sigma) \right]_{jm}$.
The latter equation holds evidently for $G_0$ as well.
The Dyson equation can then be written as
\begin{eqnarray}
G_{lj}(rr';\omega)
=
G_{0,lj}(rr';\omega)
+
\int\!\!\!\int \!\!\!dr_1dr_2\hspace{0.1cm}
G_{0,lj}(rr_1;\omega)
\Sigma_{lj}(r_1r_2;\omega)
G_{lj}(r_2 r';\omega).
\label{hfdyson3}
\end{eqnarray}
Similarly, the RPA response function and 
the residual force can be represented as
\begin{eqnarray}
R(\bvec{r}\bvec{r}';\omega)
&=&
\sum_{LM}
\frac{Y_{LM}(\hat{\bvec{r}})}{r^2}
R_L(rr';\omega)
\frac{Y^*_{LM}(\hat{\bvec{r}}')}{r'^2},
\\
\kappa(\bvec{r})
&=&
\kappa(r),
\end{eqnarray}
and consequently the self-energy can be calculated by
\begin{eqnarray}
\Sigma_{lj}(rr';\omega)
=
\sum_{l'j',L}
\frac{|\bra lj||Y_L||l'j'\ket|^2}{2j+1}
\int_{-\infty}^{\infty}\frac{d\omega'}{2\pi}
\frac{\kappa(r)}{r^2}
G_{0,l'j'}(rr';\omega-\omega')
\frac{\kappa(r')}{r'^2}
iR_L(rr';\omega').
\label{selfen2}
\end{eqnarray}

\begin{figure}
\includegraphics[width=0.8\linewidth]{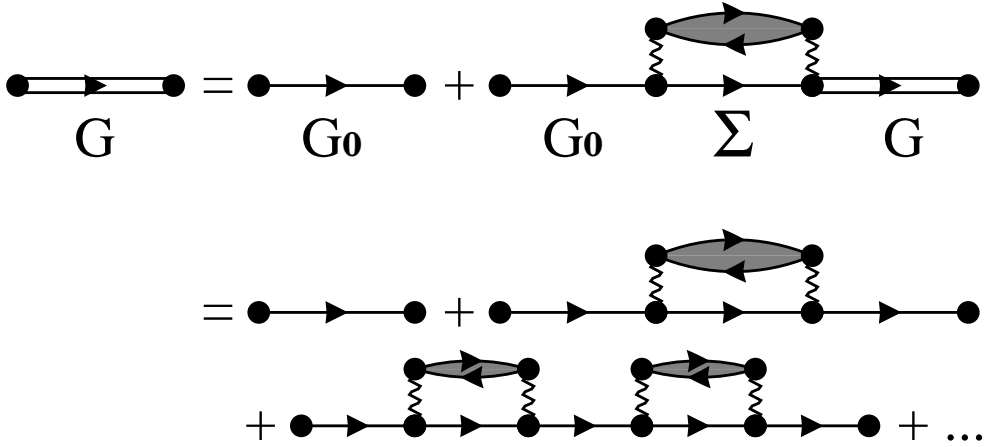}
\caption{The Feynman diagrams associated with the 
perturbative expansion of the Dyson equation [Eq. (\ref{hfdyson1}) 
or Eq. (\ref{hfdyson2})].}
\label{dysondiag}
\end{figure}

In our work, we start from the HF Green's function and
RPA response function (together with the residual force
$\kappa$) and we obtain the self-energy from 
Eq. (\ref{selfen2}). Then, we also solve numerically 
the Dyson equation in the form (\ref{hfdyson3}): for
every energy of interest this equation can be cast in
matrix form with respect to $r$ and $r'$ and solved as 
\begin{eqnarray}
G=(1-G_0\Sigma)^{-1}G_0.
\end{eqnarray}
In this way, the perturbed Green's function $G$  
contains the PVC perturbation up to infinite order, in keeping with the 
fact that it can be expressed by the Feynman diagrams 
represented in Fig. \ref{dysondiag}.

\subsection{Implementation of the proper treatment of the 
continuum}

\subsubsection{Continuum HF Green's function and 
continuum RPA response function}

As already mentioned, our goal is an implementation of
PVC that treats the continuum properly. In the case of
atomic nuclei, in particular when local functionals like
those based on the Skyrme interaction are used, considerable 
efforts have been made in this direction as far as
the HF-RPA formalism is concerned. Indeed, the Green's 
function RPA has been formulated with Skyrme forces, 
with or without \cite{Bertsch1,Bertsch2} the continuum;
the first self-consistent continuum calculations have been presented
in Ref. \cite{Liu}. In this context, proper
treatment of the continuum means that the
Schr\"odinger equation including the HF mean field
can be solved at any positive energy with the
correct boundary conditions and, based on this,
an exact representation of the HF Green's function
$G_0$ 
can be obtained.

This unperturbed HF Green's function can be written as 
\begin{eqnarray}
G_{0,lj}(rr';E)
=\frac{1}{W(u,v)}u_{lj}(r_<;E)v_{lj}(r_>;E),
\label{cGF}
\end{eqnarray}
where $u_{lj}(r;E)$ and $v_{lj}(r;E)$ are, respectively, the
regular and irregular solutions of the radial HF equation at
energy $E$, $r_>$ ($r_<$) are the larger (smaller) between
$r$ and $r'$, and
$W(u,v)$ is the Wronskian given by
\begin{eqnarray}
W(u,v;E)=\frac{\hbar^2}{2m^*(r)}\left(u_{lj}(r;E)\del{v_{lj}(r;E)}{r}-v_{lj}(r;E)\del{u_{lj}(r;E)}{r}\right).
\end{eqnarray}
$\frac{\hbar^2}{2m^*(r)}$ is the (radial-dependent) 
HF effective mass which is defined as usual, in terms of
the Skyrme force parameters, as
\begin{eqnarray}
\frac{\hbar^2}{2m^*_q(r)}
=
\frac{\hbar^2}{2m}
+
\frac{1}{4}\left\{t_1\left(1+\frac{1}{2}x_1\right)+t_2\left(1+\frac{1}{2}x_2\right)\right\}\rho(r)
-
\frac{1}{8}\left\{t_1\left(1+2x_1\right)+t_2\left(1+2x_2\right)\right\}
\rho_q(r).
\end{eqnarray}

In order to take properly into account the continuum effects 
also for the RPA phonons that lie above the threshold, 
the RPA response function appearing in the self-energy function 
will be calculated self-consistently,
using the same Skyrme Hamiltonian used to compute the mean field. 
The details of the continuum RPA calculation
have been given in previous papers \cite{sagawa,mizuyama}.
We simply recall that two-body spin orbit and Coulomb terms,
as well as spin-dependent terms, are dropped in the residual
interaction. 
It is also necessary to convert the 
continuum RPA response function into the causal function, 
because normally 
the linear response theory (RPA) is formulated in terms 
of the retarded functions. This point is further discussed 
in Appendix B.

\subsubsection{Contour integration in the complex energy plane.}
\label{sub_contour}

Formally, the equations that appear in Subsec. \ref{IIIA} are
defined in a model space which does not have an upper bound: in fact, 
the integrals over energy extend in principle from $-\infty$ to
$+\infty$, and single-nucleon as well as phonon energies
have only a natural lower bound. However, the self-energy 
function does not converge if the upper limit on 
$\omega^\prime$ [Eqs. (\ref{selfen}) and 
(\ref{selfen2})] is extended to
infinity, in keeping with the well-known ultraviolet 
divergence associated with the zero-range character of 
Skyrme forces. To avoid this, one must introduce a cutoff 
$E_{\rm cut}$.

\begin{figure}
\includegraphics[width=0.8\linewidth]{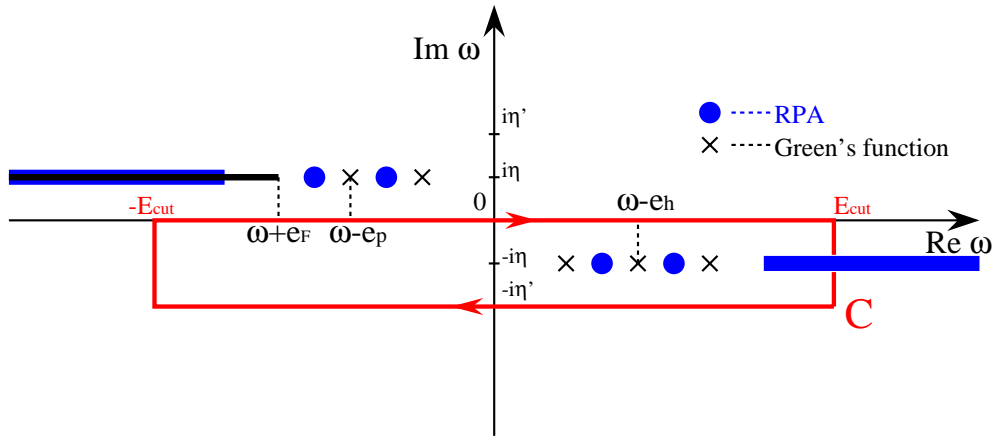}
\caption{(Color online) Contour path $C$ for the integration 
on $\omega'$ in Eq. (\ref{selfen}) and Eq. (\ref{selfen2}). 
The blue dots and lines represent the poles of the RPA response function. 
The black crosses and line represent the poles of the HF Green's 
function. For the parameters $\eta$, $\eta'$ and 
$E_{\rm cut}$ see the text.
}
\label{contour}
\end{figure}

In order to make sure that only states below that
cutoff contribute to the integrals in Eq. (\ref{selfen}) 
and (\ref{selfen2}), one can use the following procedure. 
By considering the expression (\ref{selfen2}) for the 
self-energy function, one notices that the
integral receives contribution from the poles of the  
causal HF Green's function and the causal RPA response function. 
The positions of these discrete and continuum (i.e., branch-cut) 
poles in the complex energy plane are schematically 
shown in Fig. \ref{contour}. The blue dots and line represent the 
poles of the RPA response function, while the black crosses 
and line represent the poles of the HF Green's function. 
In order to pick up correctly the contribution of the poles 
below the cutoff, we must replace the integral 
$\int_{-\infty}^{\infty}$ in Eq. (\ref{selfen2}) by 
an integral $\int_C$ over an appropriate contour path. 
We have adopted the rectangular integration path 
displayed in Fig. \ref{contour}, which is
similar to that employed in Ref. \cite{matsuo1}. It extends 
between $-E_{\rm cut}$ to $E_{\rm cut}$
on the real axis, and from 0 to -$\eta'$ on the imaginary axis.

It can also be shown that in this way one can 
reproduce the correct spectral representation 
of the self-energy function [cf. Eq. (\ref{selfenrt3})] 
in the limit of a discrete system. 

\subsubsection{Single-particle level density}\label{free}

The level density associated with the HF single-particle 
levels can be defined by using the HF Green's function $G_0$ as
\begin{eqnarray}
\rho_{0,lj}(\omega)
=
\frac{\pm 1}{\pi}\int_0^R dr\  
\mbox{Im} 
G_{0,lj}(rr,\omega),
\label{lvd}
\end{eqnarray}
where $R$ is the upper limit of the integration. For bound
states at negative energies, one is guaranteed that the result is stable with
respect to increasing $R$. 
The sign $\pm$ guarantees that the level density is positive, {\it i.e.},  
the sign +(-) refers to particle (hole) states. 
This is equivalent to the definition 
$\rho_{0,lj}(\omega)=\sum_{n}\delta(\omega-e^0_{nlj})$ 
and is normalized to 1 for bound states.
In the absence of any potential, $\rho_{0,lj}(\omega)$ reduces to the 
free particle level density $\rho_{free,lj}(\omega)$,  
obtained by replacing $G_{0,lj}$ with the Green's function for 
the free particle which satisfies $(\omega-\frac{p^2}{2m})G_{free}=1$. 
$G_{free}$ can be calculated either numerically or analytically 
using the same definition of Eq. (\ref{cGF}), but with the use 
of the wave function for the free particle. 
This free particle level density is  
$\rho_{free,lj}(\omega)
\propto\sqrt{\frac{2m}{\hbar^2}}\frac{R}{2\pi\sqrt{\omega}}$ 
for large value of $\omega$. For $R\to\infty$, $\rho_{free,lj}$ diverges.  
On the other hand, 
$\rho_{0,lj}(\omega)$ tends to $\rho_{free,lj}$ for 
large values of $\omega$. It is then useful to 
introduce a new level density $\bar{\rho}_{0,lj}(\omega)$ 
by subtracting $\rho_{free,lj}(\omega)$ \cite{shlomo}, that is, 
\begin{eqnarray}
\bar{\rho}_{0,lj}(\omega)
=
\frac{\pm 1}{\pi}\int_0^R dr\  
\mbox{Im} 
\left(
G_{0,lj}(rr,\omega)
-
G_{Free,lj}(rr,\omega).
\right)
\label{lvd2}
\end{eqnarray}
In this way, the dependence on $R$ is eliminated 
also for positive energies (cf. Fig. \ref{lvdhf}).
For $\omega<0$, there is no contribution associated 
with the free particles. As mentioned above, 
$\bar{\rho}_{0,lj}$ coincides with the usual definition 
of the level density for the single-particle levels \cite{shlomo}
\begin{eqnarray}
\label{nmlvd}
\bar{\rho}_{0,lj}(\omega)
=
\left\{
\begin{array}{ll}
\displaystyle{\sum_n\delta(\omega-\epsilon^{(0)}_{nlj})} & 
\mbox{for bound states\ }(\omega<0), \\
\displaystyle{\frac{1}{\pi}\frac{d\delta^{(0)}_{lj}}{d\omega}} & 
\mbox{for positive energy states\ }(\omega>0).
\end{array}
\right.
\end{eqnarray}

In an analogous way, we define the perturbed (HF+PVC) level 
density by using the solution of the Dyson 
equation $G_{lj}$ as
\begin{eqnarray}
\bar{\rho}_{lj}(\omega)
=
\frac{\pm 1}{\pi}\int dr\  
\mbox{Im} 
\left(
G_{lj}(rr,\omega)
-
G_{Free,lj}(rr,\omega)
\right)
\label{lvd3}
\end{eqnarray}
The peaks of the perturbed level density 
provide renormalized single-particle energies 
which include the effect of the particle-phonon 
coupling. In fact, if this coupling is small
one can expect a simple shift of the HF peaks.
Otherwise, the s.p. strength can be quite
fragmented: the associated widths 
reflect the basic decay mechanisms that are 
the nucleon decay (providing the so-called escape
width, or $\Gamma^\uparrow$) and the spreading 
into the complicated configurations made up
with nucleons and vibrations (providing the 
spreading width $\Gamma^\downarrow$). 

\section{Results}\label{results}

We shall present results for three nuclei: $^{40}$Ca, 
$^{208}$Pb and $^{24}$O. The effective Skyrme interaction 
SLy5 \cite{Chabanat} is used to calculate the HF mean
field. The calculation of the RPA response is carried out 
exactly as described in Ref. \cite{mizuyama} in the limit 
of no pairing, using all the terms of the SLy5 interaction, except  
for the two-body spin-dependent terms, the spin-orbit terms 
and the Coulomb term in the residual p-h force. 
In the calculations, the angular momentum cutoff for 
the unoccupied continuum states is set at $l_{cut}=7\hbar$ 
for $^{40}$Ca, and $l_{cut}=12\hbar$ for $^{24}$O and $^{208}$Pb,  
respectively. The radial mesh size is $\Delta r=0.2$ fm. 
The values of the parameters used in the contour integrations 
(see the discussion in \ref{sub_contour}) are $\eta =$ 0.2 MeV 
[cf. Eq. (\ref{selfen2})] and $\eta' = 2 \eta$ [cf. Eq. 
(\ref{hfdyson3})].

There are a few important issues that we wish to
stress:
\begin{enumerate}
\item
Due specifically to the zero-range character of the Skyrme 
interaction, the self-energy diverges logarithmically 
as a function of the maximum energy of the phonons, as
it has mentioned above. The first steps towards a
systematic renormalization procedure have only recently been 
started to be worked out \cite{colodiverg}. In this work, 
we shall take the usual view that the important couplings 
are those associated with the collective low-lying states
and giant resonances. We shall then include phonons 
associated with the multipolarities $2^+$, $3^-$, $4^+$ and 
$5^-$, and set an upper cutoff on the phonon energies
given by $E_{\rm cut}=60$ MeV 
because no strong peaks are present 
above this value in the calculated RPA strengths.
The way in which this cutoff is implemented has been
described in detail in \ref{sub_contour}.
\item
In the present scheme, the price to be paid for the
exact continuum treatment, is that one cannot discriminate
between the inclusion of collective and non-collective
phonons. Actually, the diagrams shown in Figs. \ref{selfendiag}\ 
and \ref{dysondiag} contain terms that violate the Pauli
principle, and these terms are larger when the phonons
are non-collective. In other words, one could expect that 
in an exact calculation the correction of the Pauli principle 
violation cancels, to a large extent, the contributions
from non-collective phonons. Although this point has never
been clarified in the available literature, to our knowledge, 
on a quantitative basis, in most of the cases the usual view
has been to take into account only the coupling to collective
states. In the work that we quoted already (the most similar
to the present one), namely in the recent calculation of 
particle-vibration coupling in $^{40}$Ca and $^{208}$Pb of 
Ref. \cite{colo10}, 
only phonons exhausting at least 5\% of the isoscalar or isovector 
non-energy-weighted sum rules have been taken 
into account. 
Therefore, we can expect that the effect of particle-vibration
coupling is larger when we calculate it with the present method,
as compared with 
Ref. \cite{colo10}. We shall come back to this issue below.
\item
The momentum-dependent part of the particle-hole interaction 
had previously been neglected in the calculation
of the particle-vibration coupling; then, in Ref. \cite{colo10} 
it was shown that its effect is important (at least in the 
case of the SLy5 interaction), and that it can be 
reasonably accounted for 
within the Landau-Migdal (LM) approximation, by 
choosing the Fermi momentum $k_F$ as 1.33 fm$^{-1}$ (that is,
at the value associated with the nuclear matter saturation density).
The LM approximation will be adopted in the following, with
the same value of $k_F$. 
\end{enumerate} 

\subsection{Results for $^{40}$Ca}

\begin{figure}[htbp]
\includegraphics[width=\textwidth,angle=-90,scale=0.6]{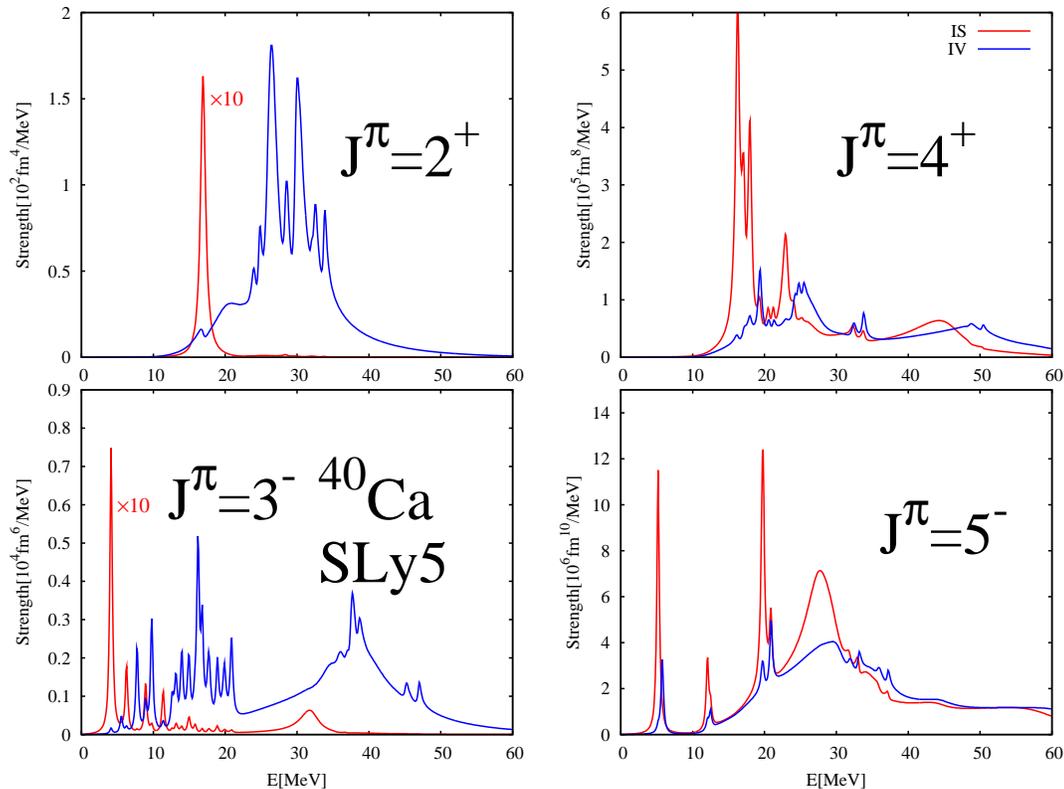}
\caption{(Color online) Isoscalar (IS) and isovector (IV) 
RPA strength functions in $^{40}$Ca
for the multipolarities 
$2^+,3^-,4^+$ and $5^-$. The IS $2^+$ 
and IS $3^-$ strength functions are reduced by a factor 10.}
\label{str_Ca40}
\end{figure}

The first essential steps of our work 
consist in the calculation of the HF spectrum and 
of the RPA strength functions.
The results are illustrated respectively in Table \ref{splvtbCa40} (HF single-particle spectrum) 
and in Fig. \ref{str_Ca40} (isoscalar and isovector RPA strength functions associated with 
the multipolarities $2^+,3^-,4^+$ and $5^-$). 
In Table \ref{lwst_Ca40} we give the theoretical
energies and transition strengths for the low-lying collective $3^-$ and $5^-$ states,
comparing them with available data.  

\begin{table}[htbp]
\renewcommand\arraystretch{1.5}
\begin{tabular}{ccccc}
\hline
\hline
Nucleus &\multicolumn{2}{c}{hole states [MeV]}\hspace{0.1cm} & \multicolumn{2}{c}{particle states [MeV]} \\[0.5mm]
\hline
$^{40}$Ca
&$1s\frac{1}{2}$ & -48.3 &  $1f\frac{7}{2}$ &-9.7\\
&$1p\frac{3}{2}$ & -35.0 &  $2p\frac{3}{2}$ &-5.3\\ 
&$1p\frac{1}{2}$ & -31.0 &  $2p\frac{1}{2}$ &-3.1\\
&$1d\frac{5}{2}$ & -22.1 &  $1f\frac{5}{2}$ &-1.3\\
&$2s\frac{1}{2}$ & -17.3 \\
&$1d\frac{3}{2}$ & -15.2 & & \\
\hline
\end{tabular}
\caption{Neutron single-particle energies in $^{40}$Ca.}
\label{splvtbCa40}
\end{table}

\begin{table}[htbp]
\renewcommand\arraystretch{1.5}
\begin{tabular}{ccccccc}
\hline
\hline
        &        & \multicolumn{2}{c}{Theory (RPA)}&& \multicolumn{2}{c}{Experiment} \\  
Nucleus & $J^\pi$ & Energy & $B(Q^{\tau=0}_J)$     && Energy     & $B(Q^{\tau=0}_J)$\\
        &        & [MeV] & [e$^2$fm$^{2J}$]        && [MeV] & [e$^2$fm$^{2J}$]\\
\cline{3-4}\cline{6-7} 
$^{40}$Ca & $3^-$  & 4.13 & $1.13\times 10^4$  && 3.74 & $1.84\times 10^4$\\
         & $5^-$  & 5.25 & $2.22\times 10^6$  && 4.49 & --\\
\hline
\end{tabular}
\caption{The theoretical values for the energy and the transition strength 
$B$ of the  
low-lying isoscalar $3^-$ and $5^-$ states in $^{40}$Ca are compared 
with the experimental data, that are taken in the case of  
the $3^-$ state from Ref. \cite{nucltableBE3} and in the case of the 
$5^-$ state from Ref. \cite{nucldata102}.}
\label{lwst_Ca40}
\end{table}

\begin{figure}[htbp]
\includegraphics[width=\textwidth,angle=-90,scale=0.6]{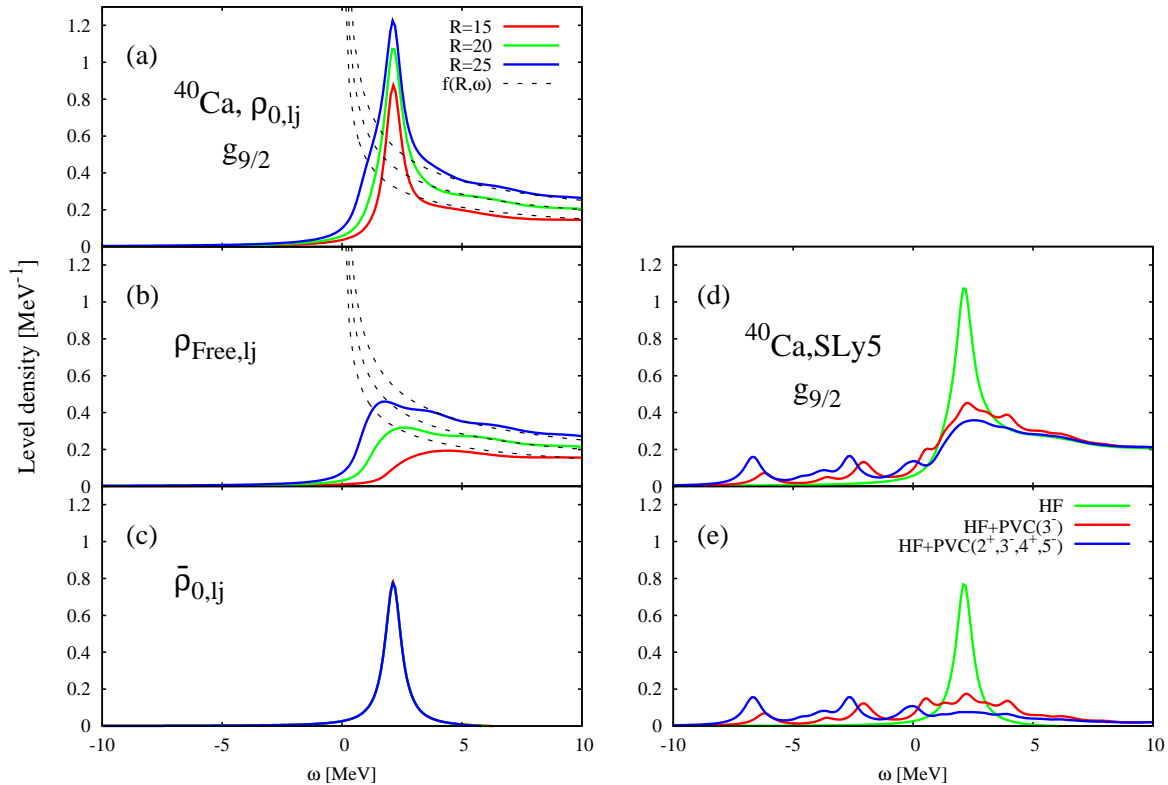}
\caption{(Color online) Neutron level density in the case of the 
g$_{9/2}$ quantum numbers for $^{40}$Ca. These quantum numbers are
associated with a single-particle resonance in the HF mean field. 
In panels (a), (b) and (c) we show respectively 
the quantities $\rho_{0,lj}$, $\rho_{Free,lj}$, 
and $\bar{\rho}_{0,lj}$ defined in 
Eqs. (\ref{lvd}-\ref{lvd3}), for three different values of the upper 
integration limit $R$, namely 15 fm (red curve), 20 fm (green curve) and 
25 fm (blue curve). The asymptotic expression
for the free particle level density, 
$f(R,\omega)=\sqrt{\frac{2m}{\hbar^2}}\frac{R}{2\pi\sqrt{\omega}}$
is shown in (a) and (b).  
Note that in (c) the curves corresponding to different values of 
$R$ practically coincide, 
because the free particle contribution is removed from $\bar{\rho}_{0,lj}$. 
In (d) we show the perturbed densities ${\rho}_{lj}$, obtained by solving 
the Dyson equation with the inclusion of the
coupling either with all multipolarities (blue curve) 
or with only the $3^-$ (red curve).
They are compared with the HF density $\rho_{0,lj}$ 
(green curve, $R$=20 fm) already shown in (a).
In (e) we show the corresponding perturbed densities $\bar{\rho}_{lj}$, 
obtained by subtracting the free-particle level density. 
They are compared with the HF density $\bar{\rho}_{0,lj}$ 
already shown in (c).}  
\label{lvdhf}
\end{figure}

In Fig. \ref{lvdhf} we compare the level densities 
$\rho_{0,lj}$, $\rho_{Free,lj}$ and 
$\bar{\rho}_{0,lj}$ defined above [cf. Eqs. (\ref{lvd}-\ref{lvd3})], in
case of neutrons and for the 
quantum number g$_{9/2}$. The results are shown for    
different values of the upper limit of integration, namely 
$R$ = 15, 20 and 25 fm. The main goal of the figure is to illustrate the effect
of the removal of the free particle level density, that 
has been formally introduced above (cf. \ref{free}).
In fact, it can be seen in panels (a) and (b) that 
for each value of $R$ the smooth tails 
of $\rho_{Free,lj}(\omega)$ and $\rho_{0,lj}$ converge,
for $\omega$ between 5 and 10 MeV, towards the 
asymptotic value $\sqrt{\frac{2m}{\hbar^2}}\frac{R}{2\pi\sqrt{\omega}}$
indicated by the dashed lines and by the symbol $f(R,\omega)$ in
the panels. 
In panel (c) we show the level density $\bar{\rho}_{0,lj}(\omega)$ 
defined by Eq. (\ref{lvd2}). The free level density is eliminated
(and the dependence on $R$ with it) so that   
one can clearly identify the pure g$_{9/2}$ resonance in the continuum. 
The inclusion of PVC leads to a quite big fragmentation of the strength,
as it can be seen in panels (d) and (e) of Fig. \ref{lvdhf}: panel
(d) is meant to mainly show that the 3$^-$ states are the most important
to produce that fragmentation, whereas in panel (e) we illustrate the 
effect of the subtraction procedure on the perturbed level density. 

\begin{figure}[htbp]
\includegraphics[width=\textwidth,angle=-90,scale=0.6]{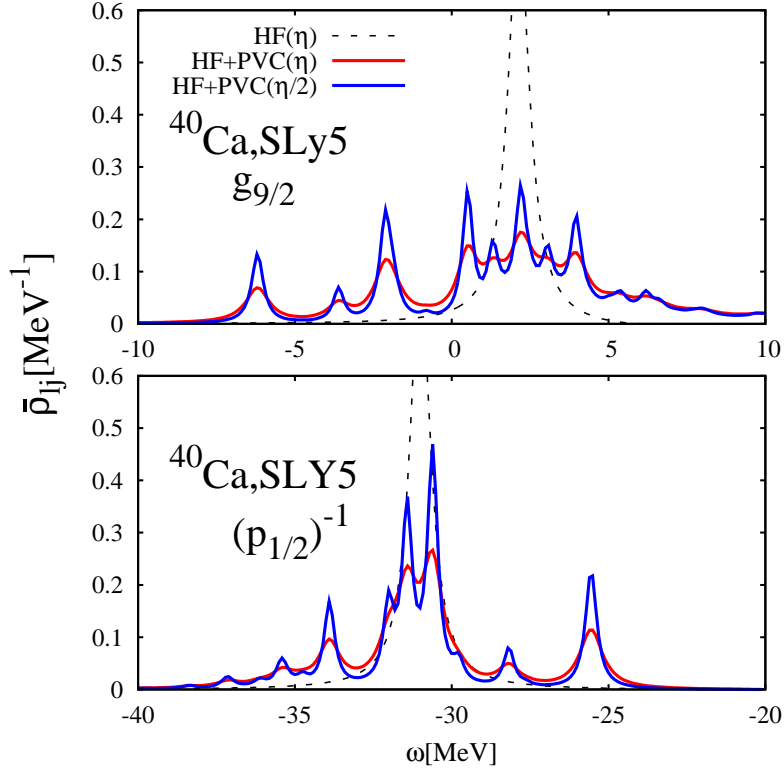}
\caption{(Color online) Dependence of the HF+PVC level density $\bar{\rho}_{lj}$
[cf. Eq. (\ref{lvd3})] on the smearing parameter $\eta$, for the
g$_{9/2}$ states (upper panel) and for the p$_{1/2}$ states (lower panel). 
Only the coupling with $3^-$ phonons is taken the account. The red  
curves show the results obtained with the standard value  $\eta=0.2$ MeV, while
the blue curves have been obtained with $\eta =$ 0.1 MeV. We also show the HF results
(dashed curves, obtained with $\eta= $ 0.2 MeV).} 
\label{lvdhf2}
\end{figure}

Since the parameter $\eta$ introduced in our definition of Green's functions
and response functions is one of 
the numerical inputs of our calculations, we have
carefully checked whether the results are sensitive to the choice of
its value. In Fig. \ref{lvdhf2}, we compare the densities $\bar{\rho}_{lj}$ 
obtained with our standard choice of the the smearing 
parameter $\eta$ = 0.2 MeV, and 
with $\eta$ = 0.1 MeV, again corresponding to $^{40}$Ca and the Skyrme set
SLy5, in the case of the quantum numbers g$_{9/2}$ and p$_{1/2}$.
It is quite reassuring that the structure displayed by the peaks of the
level density does not depend on the chosen value of $\eta$. In principle,
we may expect that a dependence of this kind shows up in the discrete
part of the spectrum and vanishes when the continuum coupling becomes
dominant: this effect can be to some extent seen in the high-energy
part (above $\approx$ 5 MeV) of the upper panel of the figure. 

\begin{figure}[htbp]
\includegraphics[width=\textwidth,angle=-90,scale=0.8]{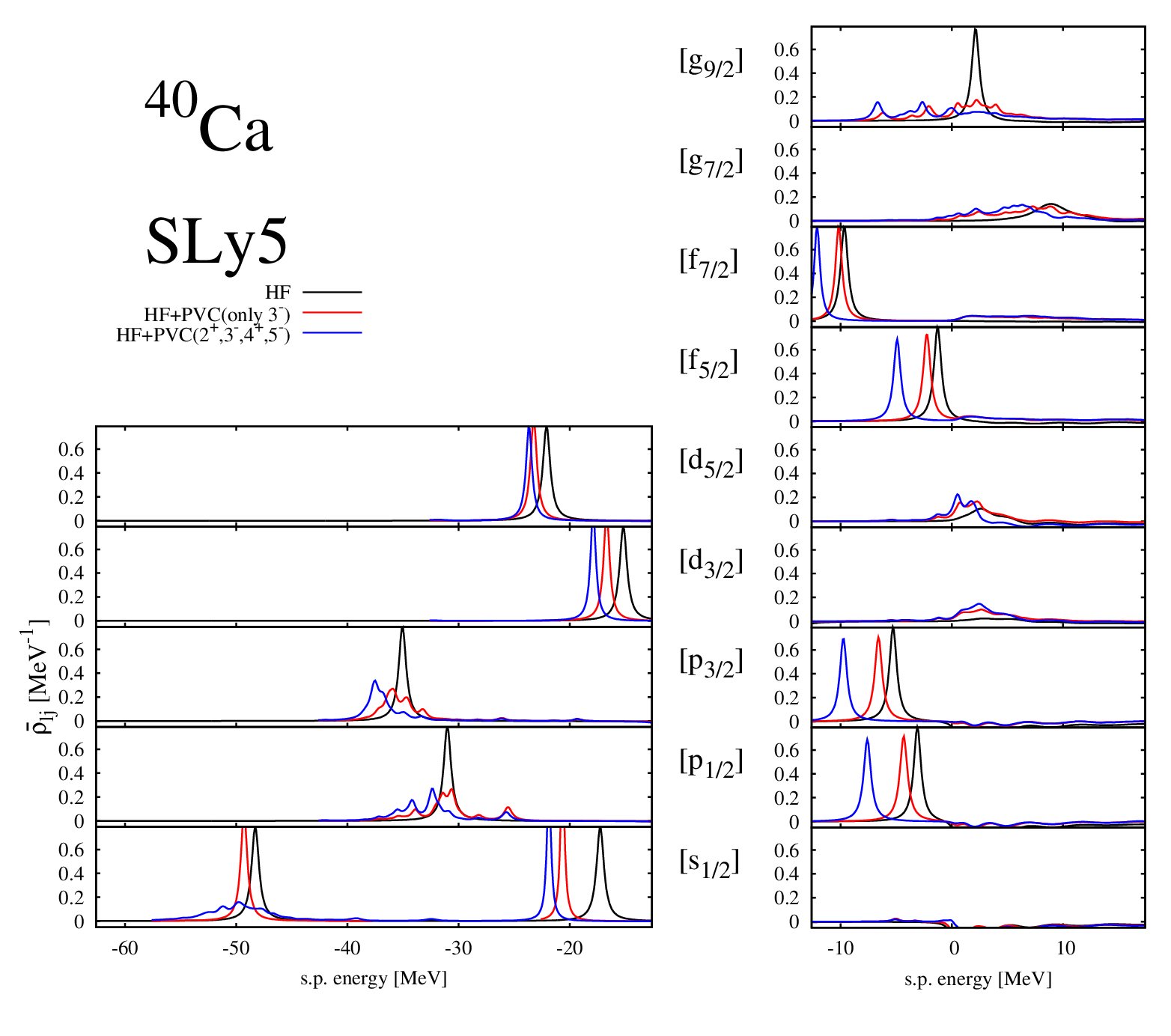}
\caption{(Color online) The single particle level density $\bar{\rho_{lj}}$ for neutron 
states in $^{40}$Ca defined by Eq. (\ref{lvd2}) for HF and Eq. (\ref{lvd3}) for HF+PVC. 
The left panel refers to hole states, the right panel to particle states. The black 
curve represents the HF level density, while the red curve and the blue curve show level 
densities resulting from the coupling to phonons. In the case of the red curve, 
only $3^-$ RPA phonons are taken into account, while in the case of the blue curve 
$2^+, 3^-, 4^+$ and $5^-$ phonons are taken into account.}
\label{lvdCa40}
\end{figure}

In Fig. \ref{lvdCa40}, we show results for the single-particle level density 
$\bar{\rho}_{lj}$ in $^{40}$Ca, associated with various quantum numbers. 
The unperturbed level density is shown by means of the black curve, 
and displays sharp peaks of equal heights at the HF energies.   
We compare in the figure the results obtained by taking into account the coupling
with all multipolarities (blue curve), or with $3^-$ phonons only (red curve).
The first qualitative remark is that for the states lying
close to the Fermi energy, both in the case of hole states
(2s$_{1/2}$, 1d$_{5/2}$ and 1d$_{3/2}$) and bound
particle states (2p$_{3/2}$, 2p$_{1/2}$, 1f$_{7/2}$ 
and 1f$_{5/2}$),
the strength remains concentrated in a single peak,
eventually acquiring a spectroscopic factor, and
the quasiparticle picture maintains its validity. This is
not true when we consider states either more far from
the Fermi surface or in the continuum, that is, at
energies where we expect the single-particle self-energy
to become larger.

The hole states are in the left part of Fig. \ref{lvdCa40}. 
For the aforementioned 2s and 1d states there is only a
shift of the HF peak. Instead, in the case of 1p$_{1/2}$ 
and 1p$_{3/2}$ states the strength is damped over a broad interval.
It is interesting to trace the origin of this fragmentation, by
restricting the summation over the phonon multipolarity and the angular 
momentum of the intermediate single-particle states in 
Eq. (\ref{selfen2}), and analyzing the contribution of 
specific configurations. 

\begin{figure}[htbp]
\includegraphics[width=\textwidth,angle=0,scale=0.5]{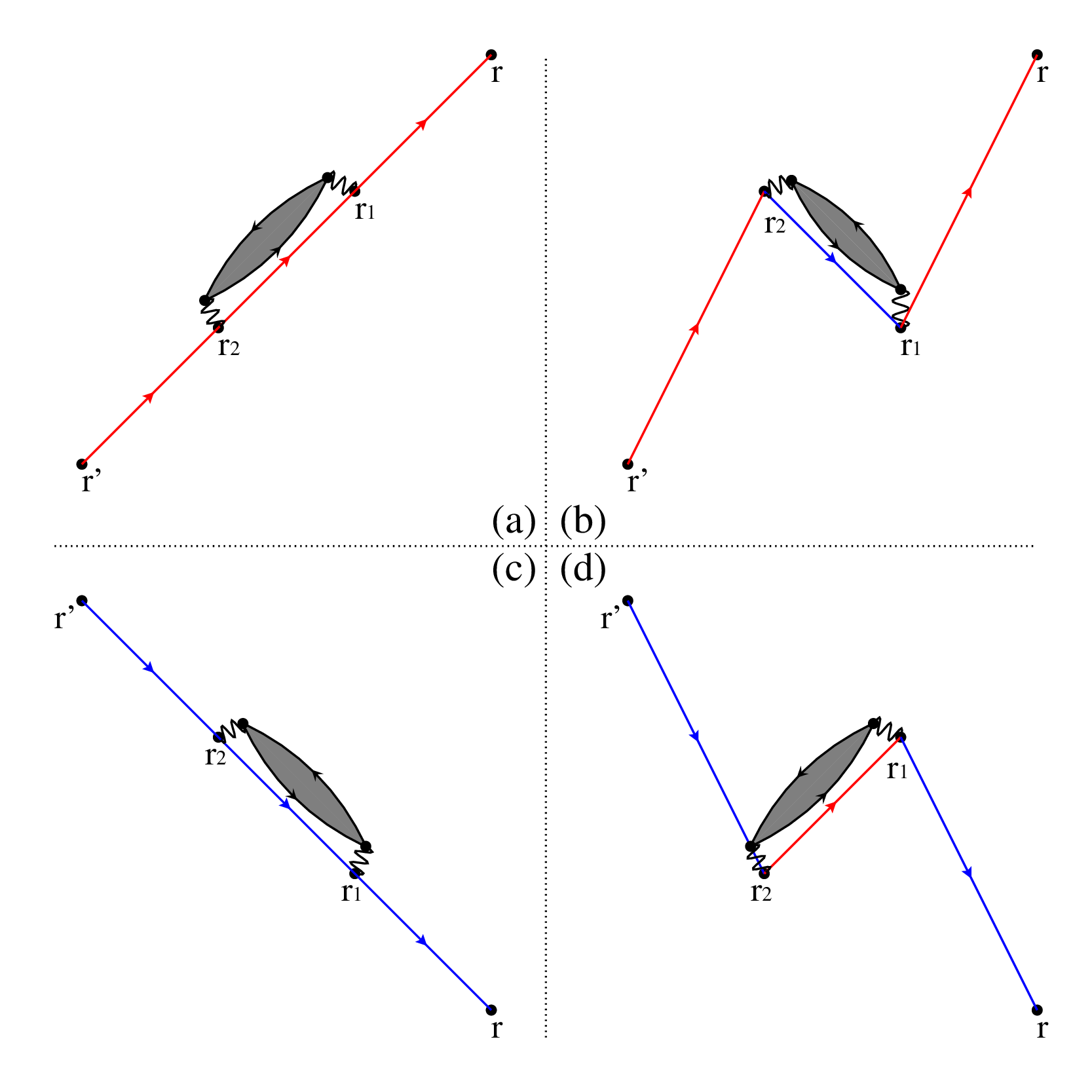}
\caption{(Color online) Feynman diagrams associated with 
particle [(a) and (b)] and hole [(c) and (d)] states. 
The shaded area denotes the RPA phonon and the wavy lines
correspond to the interaction. See the text for further 
discussion.}
\label{4_feynman_d}
\end{figure}

To help the following discussion, we depict in Fig. 
\ref{4_feynman_d} selected contributions to the particle
[(a) and (b)] and hole [(c) and (d)] self-energy. Since
our equations and numerical codes are written in the
coordinate space, these different contributions cannot,
strictly speaking, be singled out. However, at definite
energies it may happen that only one is dominant.  
We use Fig. \ref{4_feynman_d} to recall that the fragmentation of 
a hole state can only be caused by on-shell contributions 
associated with the coupling with other hole-phonon 
configurations [panel (c)], while the coupling with 
particle states [panel (d)] can only produce an 
energy shift. 

\begin{figure}[htbp]
\includegraphics[width=\textwidth,angle=-90,scale=0.6]{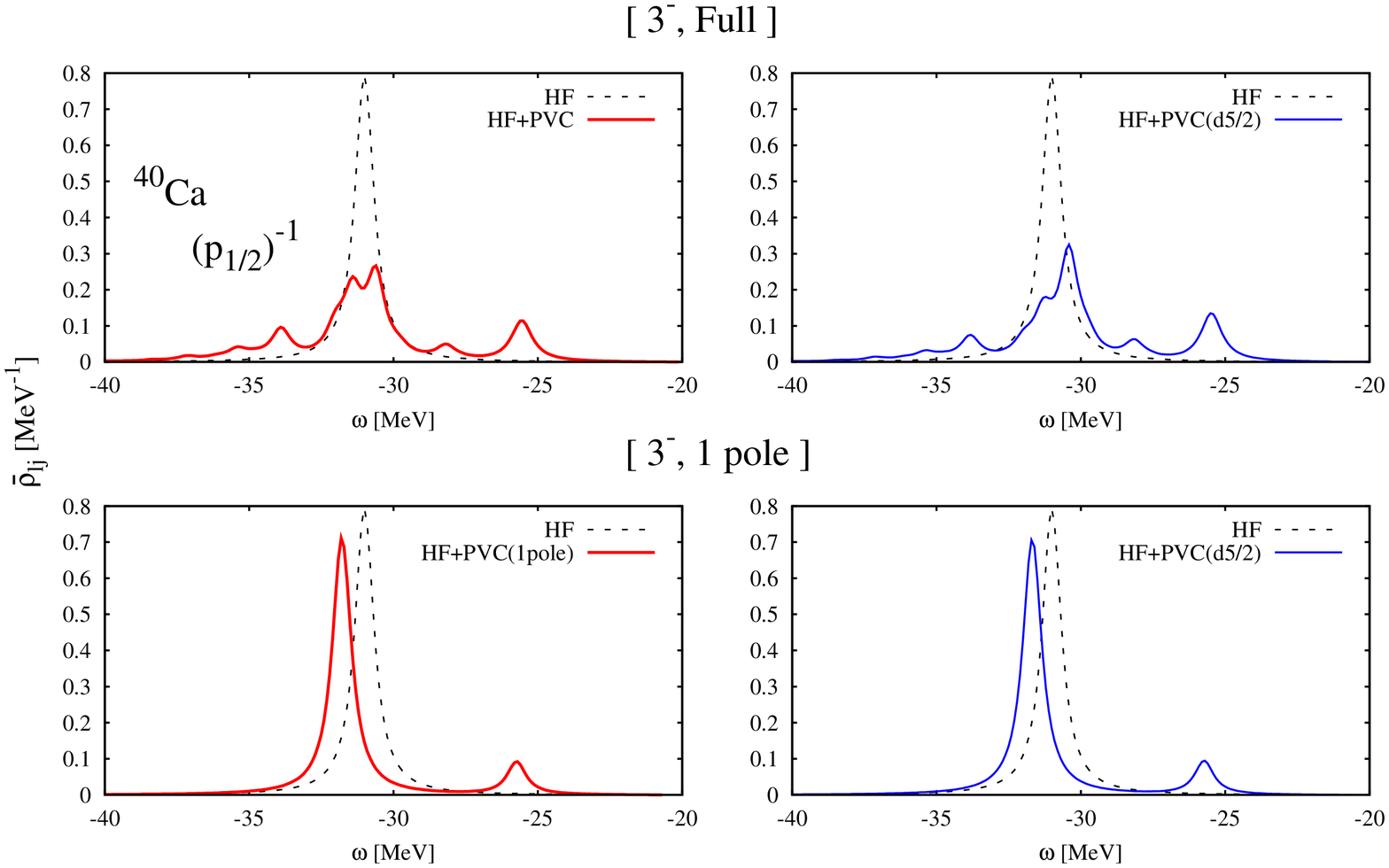}
\caption{(Color online) Analysis of the (p$_{1/2}$)$^{-1}$ 
level density in $^{40}$Ca produced by the 
coupling with $3^-$ phonons. In the upper panels, all the 
calculated RPA $3^-$ phonons 
are included in the calculation. In the 
lower panels, only the lowest $3_1^-$ phonon state is 
instead taken into account. In the left panels, 
all possible intermediate single-particle configurations 
are included. In the right panel, the intermediate 
configurations are restricted to d$_{5/2} \otimes 3^-$.}
\label{lvdhf4}
\end{figure}

The unperturbed (s$_{1/2}$)$^{-1}$ strength shows two peaks, 
associated with the 1s$_{1/2}$ ($e_{1 \rm s1/2}\approx -50$ 
MeV) and 2s$_{1/2}$ ($e_{2 \rm s1/2}
\approx -20$ MeV) single-particle states. Due to parity and angular momentum 
conservation, the only intermediate hole-phonon 
configurations s$_{1/2}$ holes can couple to 
are (d$_{3/2}$)$^{-1}\otimes 2^+$ and (d$_{5/2}$)$^{-1}\otimes 2^+$.
These coupling can lead to a strong fragmentation 
of the (1s$_{1/2}$)$^{-1}$ strength. In fact, 
the energy differences $e_{1\rm d3/2}-e_{1\rm s1/2} 
= -15.2+48.3 = 33.1$ MeV and
$e_{1\rm d5/2}-e_{1\rm s1/2}= -22.1+48.3= 26.2$ MeV 
(cf. Table \ref{splvtbCa40}) are close to
the centroid of the isovector quadrupole strength 
($E_{\rm IVGDR} = 29.2$ MeV, cf. Fig. \ref{str_Ca40}).
On the other hand, although the (1$d_{3/2}$)$^{-1}\otimes 2^+$ 
intermediate configuration could in principle 
contribute to the fragmentation of the (2s$_{1/2}$)$^{-1}$ 
state, in this case the energy 
difference is low, namely $e_{1\rm d3/2}-e_{1\rm 
s1/2}=-15.2+17.3=2.1$ 
MeV. 
In $^{40}$Ca there is no low-lying quadrupole strength, 
and therefore the HF strength is shifted but not fragmented. 
In a similar way, one concludes that the 
d$_{3/2}$ and d$_{5/2}$ hole strength, that lies close to the 
Fermi energy, cannot be fragmented.
The deeply bound p$_{1/2}$ and p$_{3/2}$ states can instead 
couple efficiently, respectively to 
the (1d$_{5/2}$)$^{-1} \otimes 3^-$ and to the 
(1d$_{3/2}$)$^{-1}\otimes 3^-$ configurations. The relevant 
energy differences lie in the range $9\sim 20$ MeV, where 
one finds substantial $3^-$ strength (cf. Fig. \ref{str_Ca40}). 
The case of $p_{1/2}$ is analyzed in more detail 
in Fig. \ref{lvdhf4}, including only $3^-$ phonons. 
It is seen that the full calculation 
(top panel, left) is very similar to the result obtained 
including only (1d$_{5/2})^{-1} \otimes 3^-$ 
configurations (top panel, right). Coupling only to the lowest 
phonons of each multipolarity (bottom panel, left) leads 
only to a modest energy shift, again caused essentially
by the (1d$_{5/2}$)$^{-1} \otimes 3_1^-$ configuration 
(bottom panel, right). The remnant of this configuration is
visible in the figures, at about -25 MeV. 

\begin{figure}[htbp]
\includegraphics[width=\textwidth,angle=-90,scale=0.6]{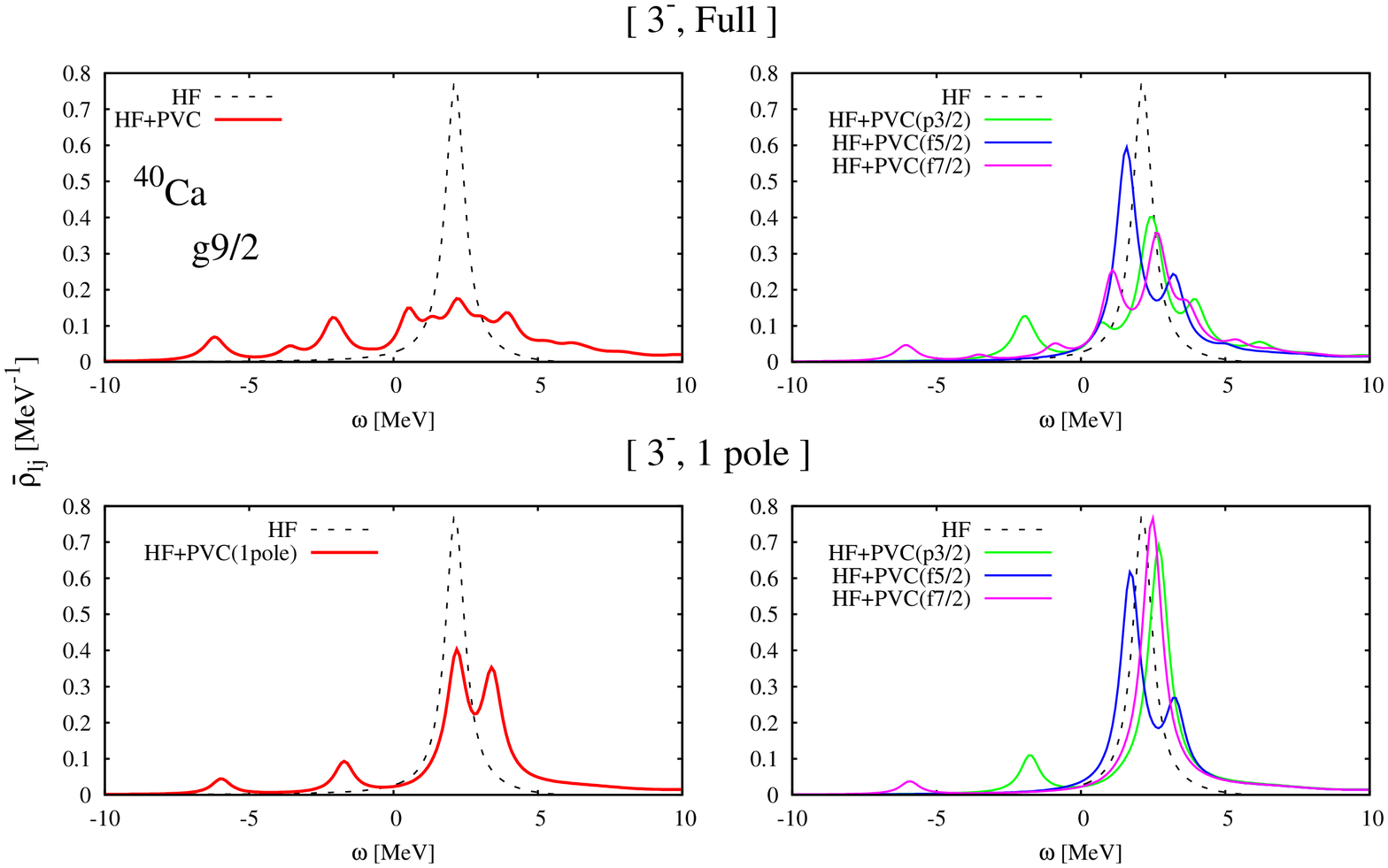}
\caption{(Color online) The same as Fig. \ref{lvdhf4} 
in the case of the g$_{9/2}$ level density.
In the right panel, we show results obtained by 
taking into account only the coupling to specific intermediate 
configurations: p$_{3/2} \otimes 3^-$ (green curve), 
f$_{5/2} \otimes 3^-$ (blue curve), 
and f$_{7/2} \otimes 3^-$ (purple curve), respectively.}
\label{lvdhf3}
\end{figure}

The results obtained for the particle states are 
depicted in the right part of Fig. \ref{lvdCa40}. 
As already mentioned, those lying close to the Fermi energy
(2p$_{3/2}$, 2p$_{1/2}$, 1f$_{7/2}$ and 1f$_{5/2}$)
can be well described within the quasiparticle picture: 
the associated single-particle strength
shows a well defined peak, which is shifted from 
the unperturbed (HF) position. The situation is quite 
different for the unbound particle states 1g$_{7/2}$ and 
1g$_{9/2}$ which are strongly fragmented. For these
states, we expect that our proper treatment of the
continuum should be particularly important.
The case of the 1g$_{9/2}$ orbital, which in the HF calculation 
is associated with a low-lying resonance lying at 
about 2.5 MeV, is analyzed in more detail in Fig. \ref{lvdhf3}. 
We consider only the coupling with $3^-$ phonons, since
they produce most of the fragmentation. By comparing 
the left and the right top panels of Fig. \ref{lvdhf3}, 
one concludes that the strong fragmentation of the 
resonant level is caused by the coupling with several intermediate 
configurations [namely (p$_{3/2}$)$\otimes 3^-$, 
(f$_{5/2}$)$\otimes 3^-$ and (f$_{7/2}$)$\otimes 3^-$]. 
The two satellite peaks found at $\approx -6$ MeV and
at $\approx -2$ MeV are produced by specific configurations, 
associated with the lowest $3^-_1$ collective state
(cf. bottom panel, right). This phonon is responsible 
for about half of the total width (compare bottom panel, 
left with top panel, left).    

\begin{figure}[htbp]
\includegraphics[width=\textwidth,angle=0,scale=0.3]{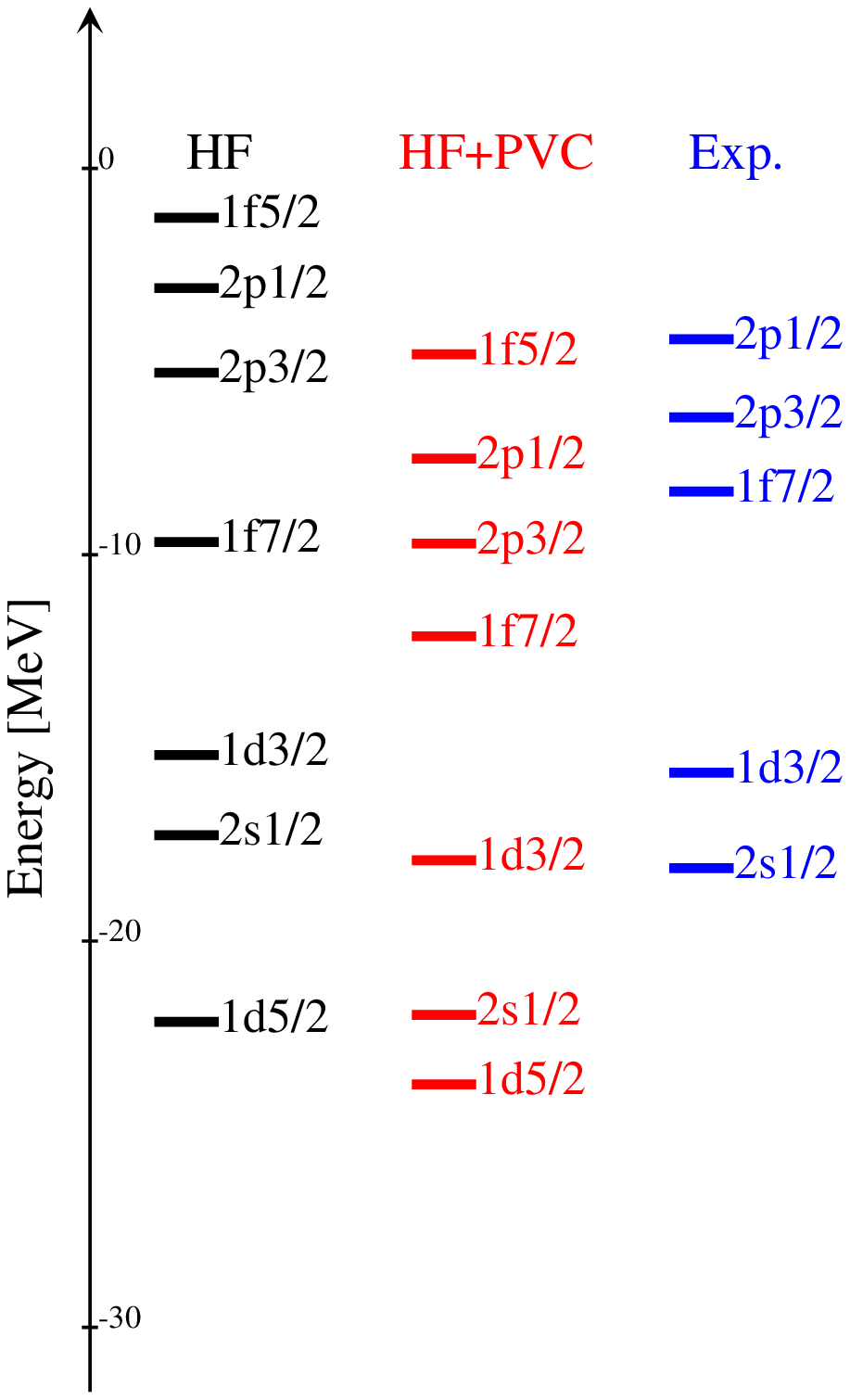}
\caption{(Color online) Energy of the seven 
neutron bound states lying close to the Fermi energy in $^{40}$Ca, 
calculated in HF (left) 
and by taking into account the coupling to phonons (right).}
\label{splv_Ca40}
\end{figure}

In Fig. \ref{splv_Ca40} we compare the position of the 
seven HF energy levels lying close to the Fermi energy 
with the position of the shifted levels, 
deduced from Fig. \ref{lvdCa40}. Only for these levels
a centroid energy is quite meaningful because, as we
have already emphasized, for these levels essentially only
one peak exists when one looks at the PVC results and
the quasiparticle picture holds. Also for these levels, and
for them only, one can attempt a comparison with the
results of Ref. \cite{colo10} that have been obtained
through second-order perturbation theory and in a very
similar scheme. The results are in overall agreement with those of 
Ref. \cite{colo10}, although the magnitude of the present 
energy shifts is larger. In fact, while in Ref. \cite{colo10} 
the shifts are typically between -1 and -2 MeV, here
they range between -1.5 and -4.5 MeV. We attribute this
difference mainly to the coupling with non collective 
phonons. As it was discussed alreday in Ref. \cite{colo10}, 
the energy shifts are mostly due to coupling with intermediate 
configurations including an octupole phonon. If we compare
the theoretical results with experiment, we must probably conclude
that a re-fitting of the effective force (SLy5 in the present case) 
is needed if this has to be used outside the mean-field framework. 
In fact, the HF-PVC results need a global upward shift in energy.

\subsubsection{Comparison with the experimental data in $^{40}$Ca}

\begin{figure}[htbp]
\includegraphics[width=\textwidth,angle=-90,scale=0.8]{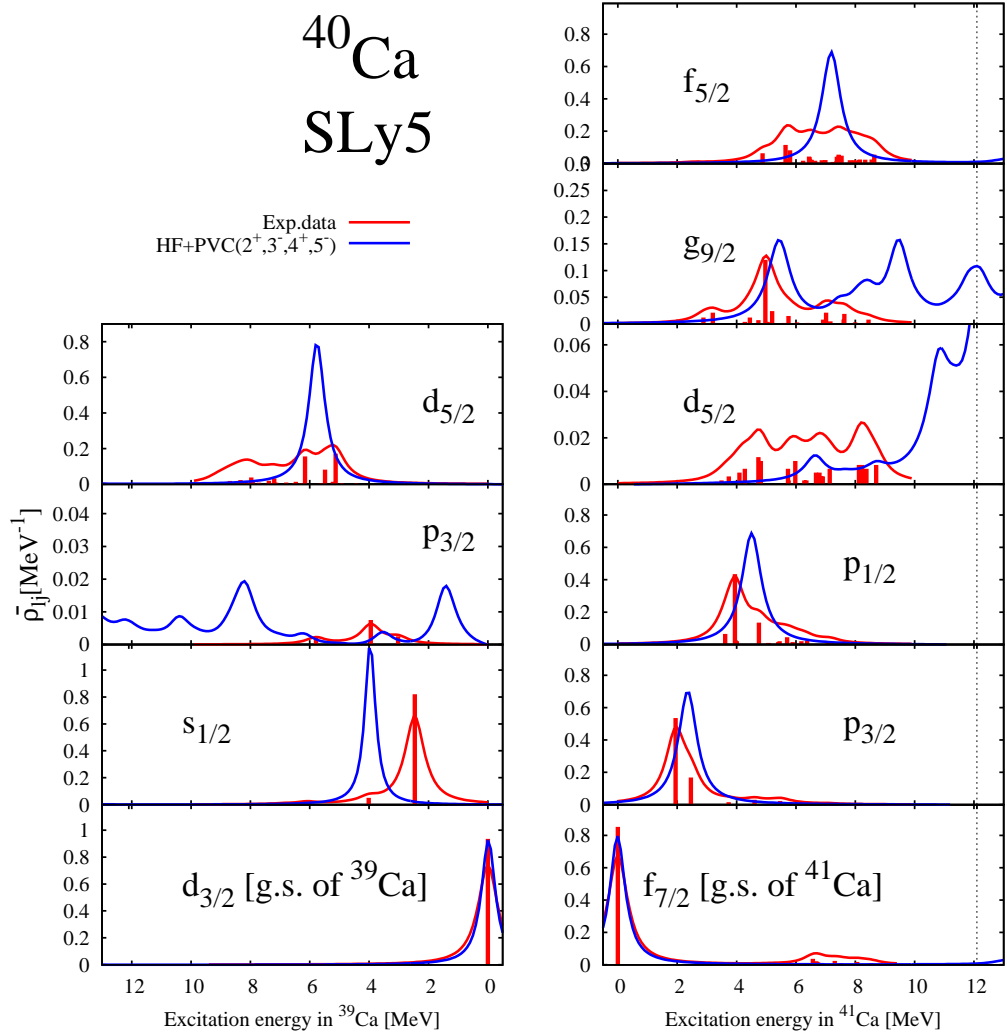}
\caption{(Color online) The theoretical level densities $\bar{\rho_{lj}}$ 
(blue curves) are shown as a function of the excitation energy in $^{39}$Ca 
(hole states, left panels) and in $^{41}$Ca (particle states, right panels).
Except for the change of scale, the results are the same already shown in 
Fig. \ref{lvdCa40}.
They are compared with the experimental spectroscopic factors taken from 
\cite{nucldataA39,nucldataA41}: these are represented in histogram form
and also convoluted with Lorentzian functions of width 0.4 MeV (red curve). 
The vertical dotted line shows the one-neutron theoretical threshold 
energy obtained within HF+PVC (this energy is then the 
continuum threshold measured by setting at zero the energy 
of the renormalized 1f$_{7/2}$ state already
displayed in Figs. \ref{lvdCa40} and \ref{splv_Ca40}).
}
\label{compexp_Ca40}
\end{figure}

\begin{figure}[htbp]
\includegraphics[width=\textwidth,angle=-90,scale=0.8]{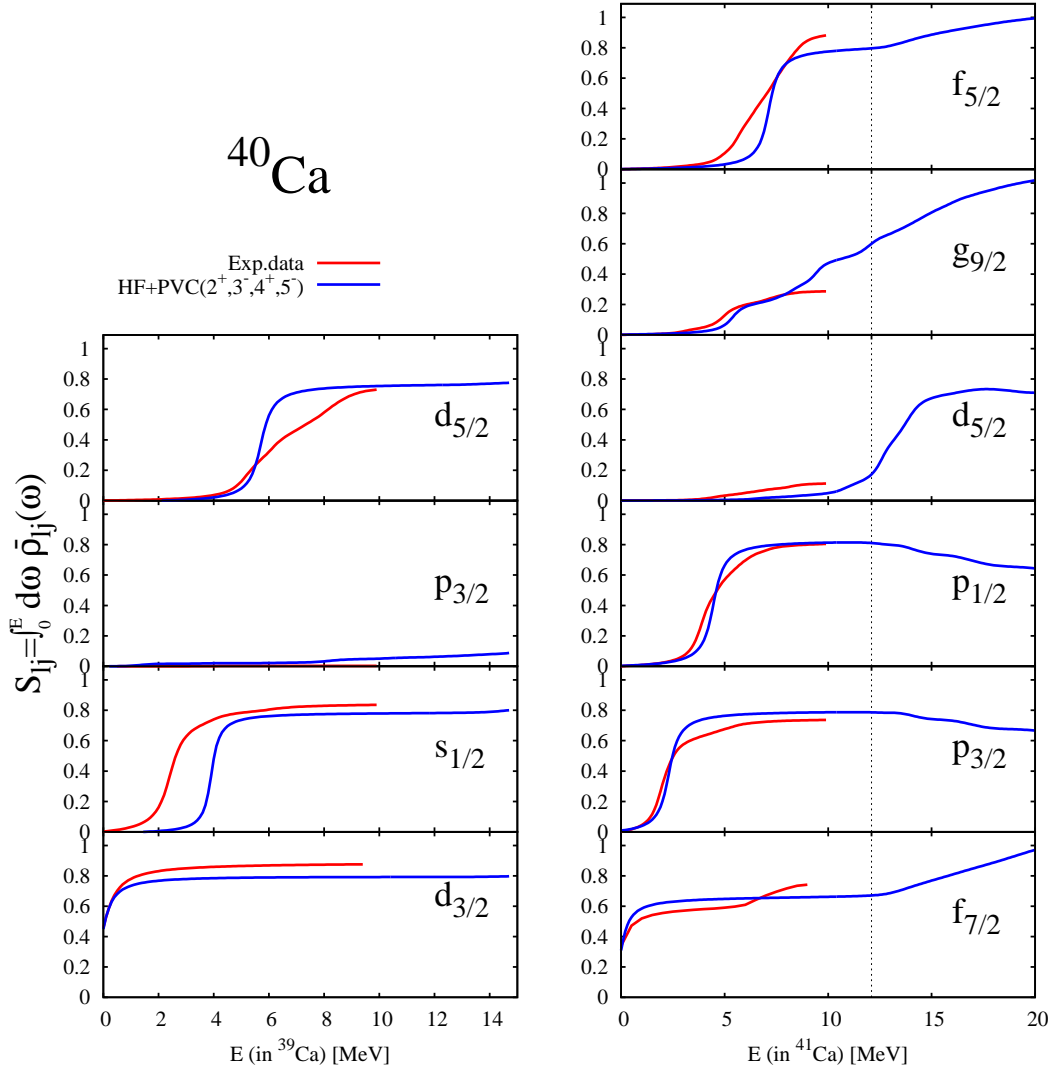}
\caption{(Color online) The integrated level density as a function 
of excitation energy for $^{39}$Ca (left panel) and $^{41}$Ca (right 
panel), respectively. The red curves correspond to the experimental data, 
while the blue curves are our HF+PVC calculation.}
\label{compexp_Ca40_sum}
\end{figure}

\begin{table}[t!]
\renewcommand\arraystretch{1.5}
\begin{tabular}{ccccccc}
\hline
\multicolumn{7}{c}{$^{40}$Ca} \\
\hline
\multicolumn{3}{c}{Holes} & & \multicolumn{3}{c}{Particles} \\
& \multicolumn{2}{c}{$S_{lj}(^{39}$Ca)} & &  & \multicolumn{2}{c}{$S_{lj}(^{41}$Ca)} \\
\cline{2-3}\cline{6-7}
$J^\pi$  & Exp.                & Theory & & $J^\pi$ & Exp. & Theory \\
d$_{3/2}$ & 0.88                & 0.80 & & f$_{7/2}$ & 0.74   & 0.66 \\
s$_{1/2}$ & 0.84                & 0.80  & & p$_{1/2}$ & 0.80 & 0.81 \\
p$_{3/2}$ & $2.9\times 10^{-3}$  & 0.05  & & p$_{3/2}$ & 0.73 & 0.79 \\
d$_{5/2}$ & 0.73 & 0.75                 & & d$_{5/2}$ & 0.11 & 0.04 \\
&  &                                    & & f$_{5/2}$ & 0.88 & 0.77 \\
&  &                                    & & g$_{9/2}$ & 0.28 & 0.36 \\
\hline
\end{tabular}
\caption{Experimental spectroscopic factors $S_{lj}$ obtained from one-nucleon
transfer reactions for hole and particle states in $^{39}$Ca and $^{41}$Ca,
compared to the integral of the theoretical level density performed up to
an excitation energy of 10 MeV (cf. Fig. \ref{compexp_Ca40_sum}).}
\label{num_compexp_Ca40_sum}
\end{table}

The experimental single-particle strength of $^{40}$Ca is obtained 
from $^{40}\mbox{Ca}(\rm p,\rm d)^{39}\rm Ca$ pickup 
(for hole states) and $^{40}\mbox{Ca}(\rm d,\rm p)^{41}\rm Ca$ stripping 
reactions (for particle states), by comparing 
the measured cross sections with Distorted Wave Born Approximation (DWBA) 
calculations performed with conventional assumptions. 
In particular, one usually assumes that the wavefunction of the transferred 
nucleon, $\phi_{nlj}$ can be taken as an eigenfunction of a static mean 
field potential, by adjusting the depth of that potential so
that the binding energy becomes equal to the experimental separation 
energy and the correct asymptotic dependence is guaranteed. The comparison 
with the level density obtained in a calculation like the present one, 
although not straightforward, is reasonable for levels which are
well described by the one-quasiparticle approximation. 
In the previous subsection, we have seen that this is indeed
the case for several states close to the Fermi energy:  
for them, the single-peak associated with a definite value
of the number of nodes $n$, appearing in HF, persists. A diagonal,
even perturbative, approximation for the mass operator is quite
appropriate.
However, for states characterized by a broad distribution in energy,
when several values of $n$ are mixed, the comparison with a simple 
DWBA calculation is likely to be less reliable (cf., e.g., the discussion 
in Ref. \cite{Satchler}).
In principle, one should rather perform a direct theoretical calculation 
of the transfer cross section, using the wavefunctions that include 
many-body correlations. This goes beyond the scope of the current 
paper, and in the following we shall limit ourselves to a simple comparison 
with the spectroscopic factors reported in the experimental 
papers \cite{nucldataA39,nucldataA41}. 
Our results are comparable to those obtained in Ref. \cite{eckle}, where the
distribution of single-particle strength in $^{40}$Ca was calculated in a
(discrete) quasiparticle-coupling model going beyond the diagonal approximation.  
The red histogram bars in Fig. \ref{compexp_Ca40} show the experimentally 
determined spectroscopic factors, which are convoluted with 
Lorentzian functions having a width equal to 0.4 MeV
to produce the red continuous lines. These can be compared with our 
theoretical level densities (blue continuous lines). 
The dotted vertical line shows the calculated threshold 
for one-neutron emission, 
which overestimates the experimental value by about 2 MeV. 
 
The total single-particle strength associated with the different 
quantum numbers ${l,j}$, obtained by integrating the level densities 
displayed in Fig. \ref{compexp_Ca40} up to $\rm E= 10$ MeV is reported 
in Table \ref{num_compexp_Ca40_sum}. One finds an overall satisfactory agreement
between theory and experiment. The position of the centroid energies is 
reasonably well reproduced, except that in the case of the s$_{1/2}$
strength, where the theoretical centroid energy is too low by about 2 MeV. 
In general, theory tends still to underestimate the fragmentation
of the single-particle strength: this occurs in particular for the d$_{5/2}$ 
strength (for particles and holes), and for the f$_{5/2}$
strength (for particles). This can be seen 
also from Fig. \ref{compexp_Ca40_sum}, 
where we show the cumulated experimental and theoretical 
strength distributions. The latter distributions 
tend to show a sharper increase. 
This can be attributed to several reasons. Among the possible ones, we point
out that 
the present RPA calculation underestimates the experimental value of the 
collectivity of the low-lying 3$^-_1$ phonon (cf. Table \ref{lwst_Ca40}) 
and cannnot describe in detail the ISGQR strength distribution. If one 
could include   
its admixture with two particle-two hole configurations, these could 
shift part of the strength at lower energy and increase the effect of
the coupling with the single-particle strength. A more fragmented ISGQR distribution
would be in better agreement with experiment \cite{Kamer,Harakeh}; however,
such a calculation would require to go beyond the formalism of the
present work.

\subsection{Results for $^{208}$Pb}

\begin{figure}[htbp]
\includegraphics[width=\textwidth,angle=-90,scale=0.7]{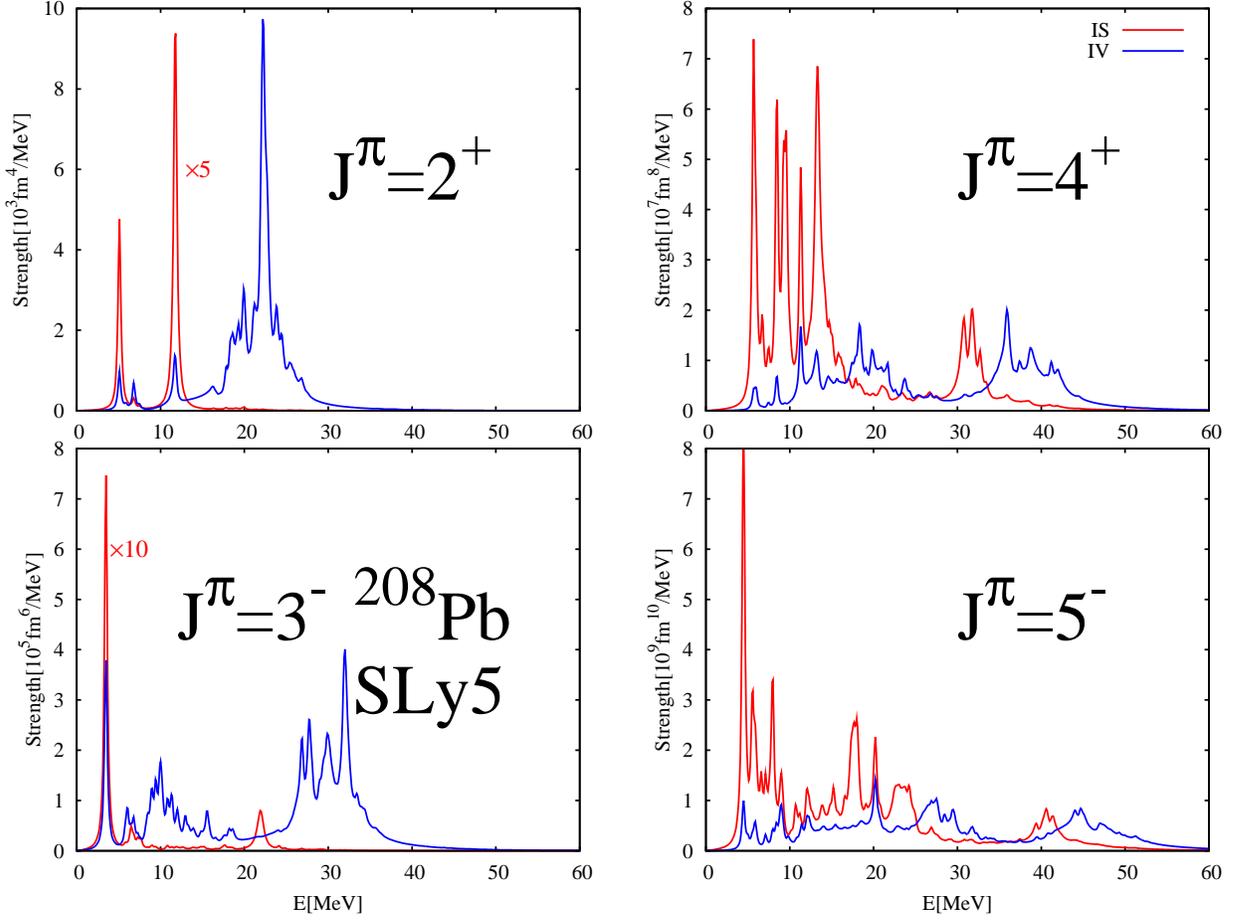}
\caption{(Color online) The RPA strength functions in $^{208}$Pb 
obtained from our RPA calculations. 
The IS $2^+$ strength is reduced by a factor 5, while 
the IS $3^-$ strength is reduced by a factor 10.}
\label{str_Pb208}
\end{figure}

\begin{table}[htbp]
\renewcommand\arraystretch{1.5}
\begin{tabular}{ccccccc}
\hline
\hline
&        & \multicolumn{2}{c}{Theory (RPA)}&& 
\multicolumn{2}{c}{Experiment} \\  
Nucleus & $J^\pi$ & Energy & $B(Q^{\tau=0}_J)$     
&& Energy     & $B(Q^{\tau=0}_J)$\\
         &        & [MeV] & [e$^2$fm$^{2J}$]        && [MeV] 
& [e$^2$fm$^{2J}$]\\
\cline{3-4}\cline{6-7} 
$^{208}$Pb 
& $2^+$  &  5.12 & $2.35\times 10^3$  && 4.09 & $3.00\times 10^3$\\
& $3^-$  &  3.49 & $7.08\times 10^5$  && 2.62 & $6.11\times 10^5$\\
& $4^+$  &  5.69 & $5.16\times 10^6$  && 4.32 & $15.5\times 10^6$\\
& $5^-$  &  4.49 & $4.96\times 10^8$  && 3.20 & $4.47\times 10^8$\\
\hline
\end{tabular}
\caption{Energies and electromagnetic transition 
probabilities associated with the low-lying 
isoscalar
collective states in
$^{208}$Pb. The experimental data are taken from 
Refs. \cite{nucltableBE2,nucltableBE3}.}
\label{lwst_Pb208}
\end{table}

\begin{figure}[htbp]
\includegraphics[width=\textwidth,angle=-90,scale=1.2]{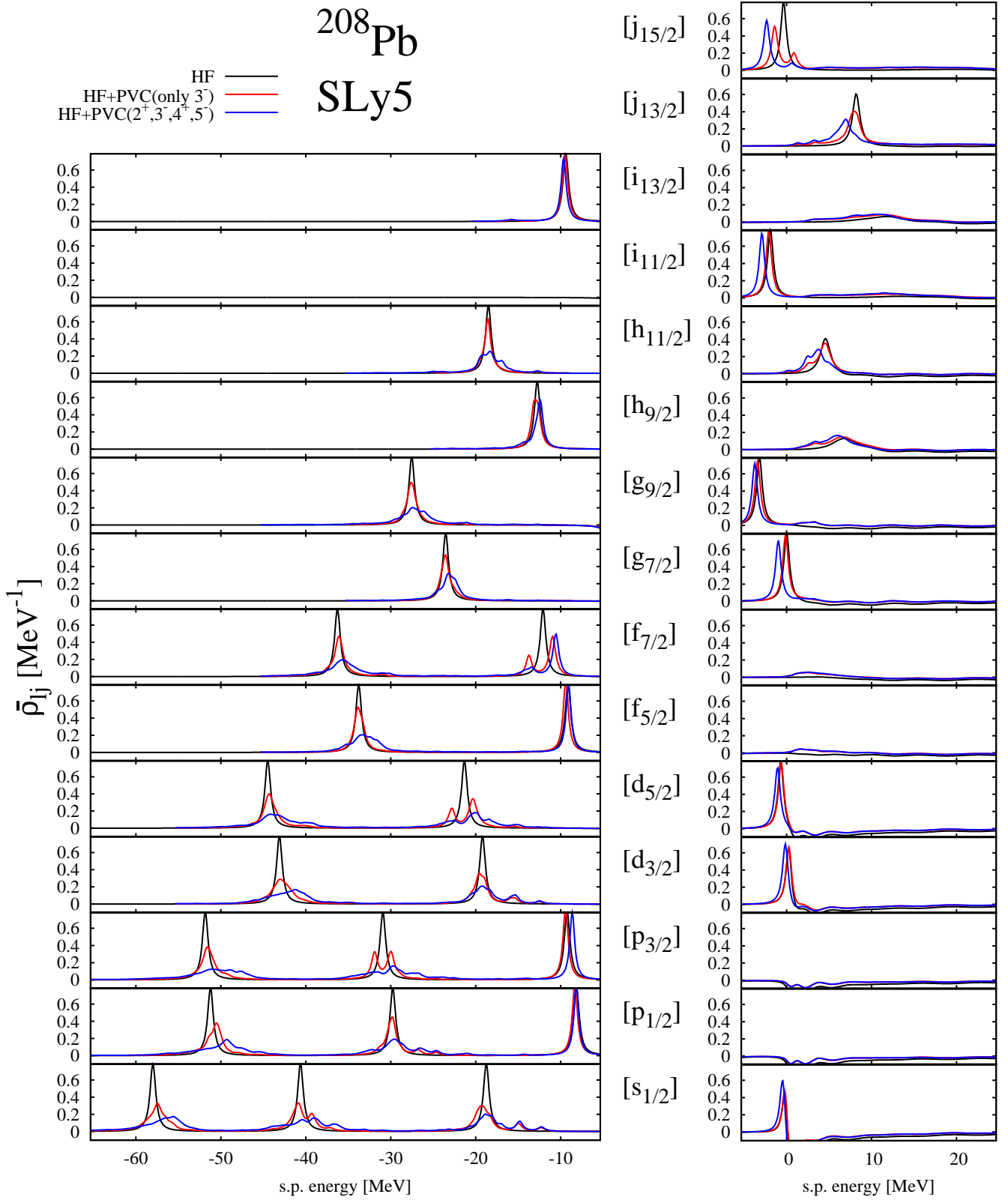}
\caption{(Color online) The same as Fig. \ref{lvdCa40} in 
the case of $^{208}$Pb.}
\label{lvdPb208}
\end{figure}

In Fig. \ref{str_Pb208} 
we provide an overall view of the calculated RPA multipole strength in 
$^{208}$Pb, while the energy and transition strength 
of the lowest states of each multipolarity are reported in Table \ref{lwst_Pb208}.
The properties of the low-lying states are
reproduced reasonably well by our calculation, 
with the partial
exception of the transition probability associated
with the 4$^+$ state.
In Fig. \ref{lvdPb208}, we show the results of our systematic 
calculation of the level densities, displaying 
the outcome of the full HF+PVC 
calculation including 2$^+$, 3$^-$, 4$^+$ and 5$^-$ phonons (blue curve), 
as well as the results obtained by including only the 
3$^-$ phonons (red curve), in comparison with the 
HF results (black curve).

\begin{table}[htbp]
\renewcommand\arraystretch{1.5}
\begin{tabular}{ccccc}
\hline
\hline
Nucleus &\multicolumn{2}{c}{Hole states [MeV]}\hspace{0.1cm} & 
\multicolumn{2}{c}{Particle states [MeV]} \\[0.5mm]
\hline
$^{208}$Pb
  &$1\rm s\frac{1}{2}$ &  -58.0 &  $4\rm s\frac{1}{2}$  &   -0.1 \\
  &$2\rm s\frac{1}{2}$ &  -40.6 &  $3\rm d\frac{5}{2}$  &   -0.7 \\
  &$3\rm s\frac{1}{2}$ &  -18.8 &  $2\rm g\frac{9}{2}$  &   -3.2 \\
  &$1\rm p\frac{1}{2}$ &  -51.2 &  $1\rm i\frac{11}{2}$ &   -1.9 \\
  &$2\rm p\frac{1}{2}$ &  -29.8 &  $1\rm j\frac{15}{2}$ &   -0.4 \\
  &$3\rm p\frac{1}{2}$ &   -8.1 & & \\
  &$1\rm p\frac{3}{2}$ &  -51.8 & & \\
  &$2\rm p\frac{3}{2}$ &  -30.9 & & \\
  &$3\rm p\frac{3}{2}$ &   -9.2 & & \\
  &$1\rm d\frac{3}{2}$ &  -43.1 & & \\
  &$2\rm d\frac{3}{2}$ &  -19.2 & & \\
  &$1\rm d\frac{5}{2}$ &  -44.5 & & \\
  &$2\rm d\frac{5}{2}$ &  -21.3 & & \\
  &$1\rm f\frac{5}{2}$ &  -33.8 & & \\
  &$2\rm f\frac{5}{2}$ &   -9.1 & & \\
  &$1\rm f\frac{7}{2}$ &  -36.3 & & \\
  &$2\rm f\frac{7}{2}$ &  -12.1 & & \\
  &$1\rm g\frac{7}{2}$ &  -23.5 & & \\
  &$1\rm g\frac{9}{2}$ &  -27.5 & & \\
  &$1\rm h\frac{9}{2}$ &  -12.8 & & \\
  &$1\rm h\frac{11}{2}$&  -18.5 & & \\
  &$1\rm j\frac{13}{2}$&   -9.4 & & \\
\hline
\end{tabular}
\caption{Skyrme Hartree-Fock single-particle energies for $^{208}$Pb 
obtained with the force SLy5. Only states at negative energy are reported  
in the Table.}
\label{splvtbPb208}
\end{table}

\begin{figure}[htbp]
\begin{minipage}{0.45\linewidth}
\includegraphics[width=\textwidth,angle=0,scale=0.8]{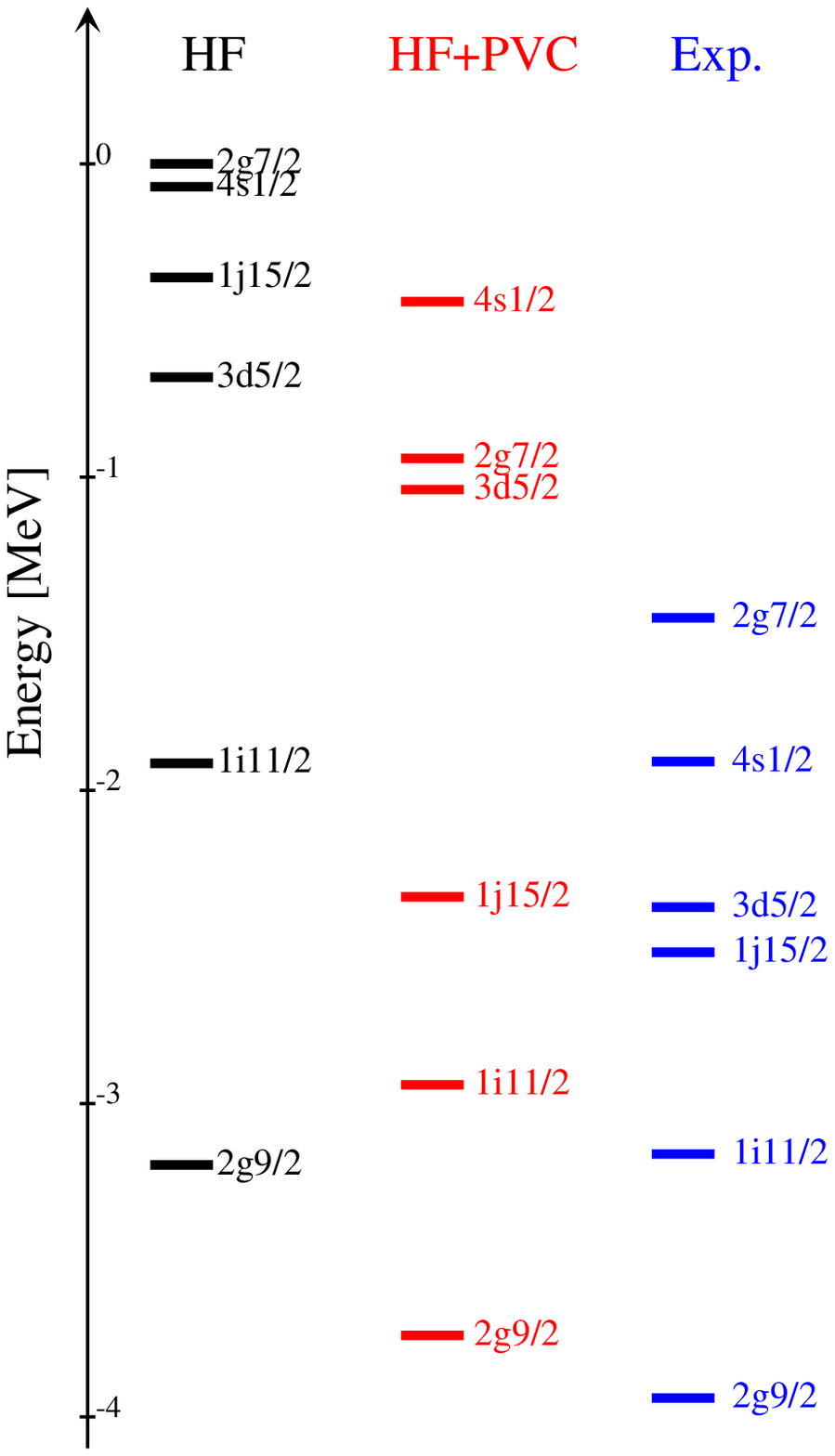}
\caption{(Color online) Neutron particle states in $^{208}$Pb.}
\label{splv_Pb208_p}
\end{minipage}
\begin{minipage}{0.45\linewidth}
\includegraphics[width=\textwidth,angle=0,scale=0.7]{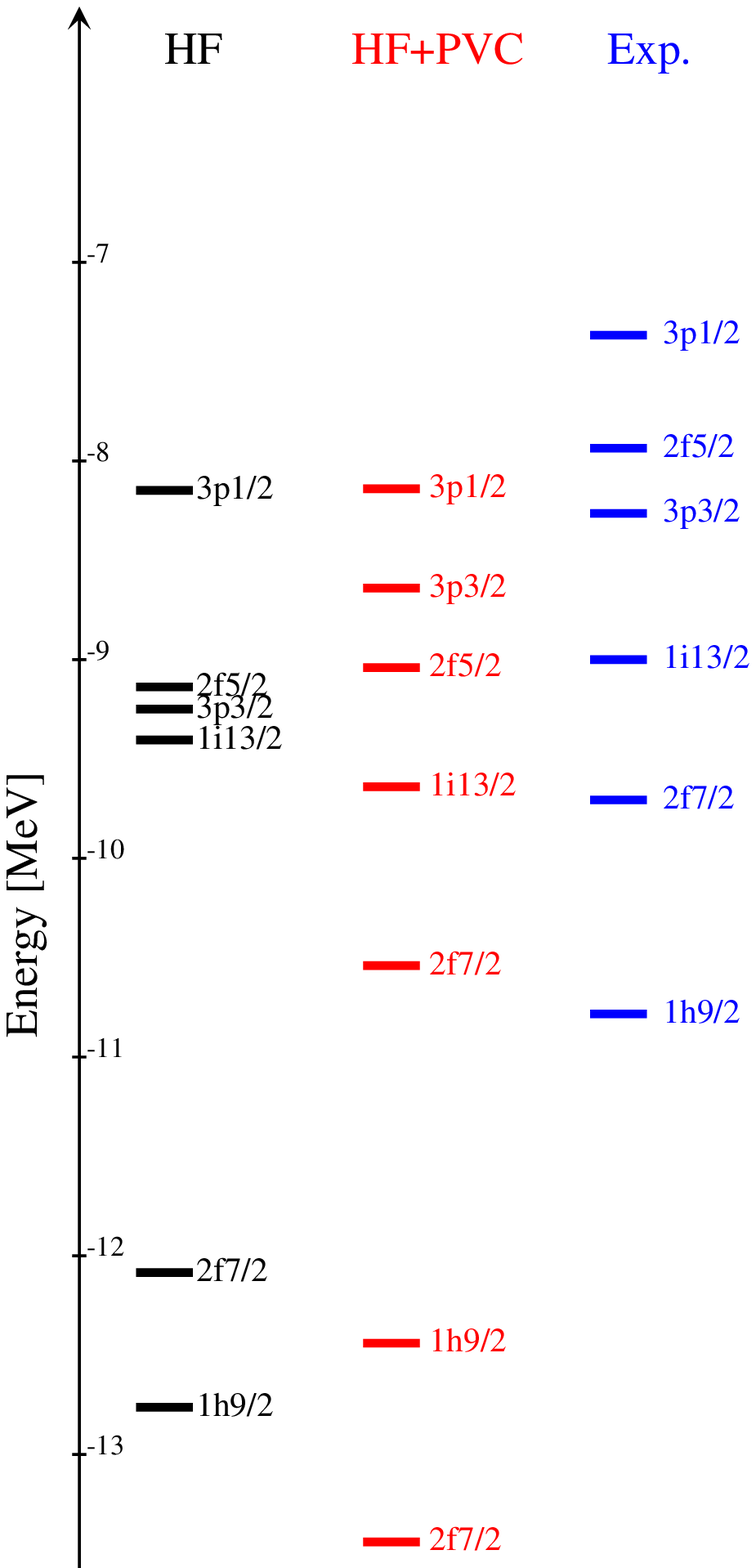}
\caption{(Color online) Neutron hole states in $^{208}$Pb.}

\label{splv_Pb208_h}
\end{minipage}
\end{figure}

The HF single-particle spectrum calculated for $^{208}$Pb is reported in Table \ref{splvtbPb208}
and illustrated in the left column of Fig. \ref{splv_Pb208_p} and Fig. \ref{splv_Pb208_h}. 
In the central column of  Fig. \ref{splv_Pb208_p} and Fig. \ref{splv_Pb208_h}
we show the position of the main peaks obtained from the full HF+PVC 
calculation, whereas in the right column we include the experimental results.
From an overall look at Fig. \ref{lvdPb208}, we can notice that 
the quasiparticle picture
(a single peak emerging from the PVC calculation,
with shifted energy and renormalized integral with respect
to HF) holds for most of the valence hole states, that is,
for 3p$_{1/2}$, 2f$_{5/2}$, 3p$_{3/2}$, 1i$_{13/2}$ and 
1h$_{9/2}$. A partial exception is constituted by the 
state 2f$_{7/2}$, that acquires a double structure mainly due
to the coupling with the 
3$^-_1 \otimes \rm i_{13/2}$ configuration. The inclusion of 
PVC brings the 
relative position of the valence hole states in much better agreement 
with experiment. 
For particle states, the quasi-particle picture
seems to be valid for 2g$_{9/2}$, 1i$_{11/2}$, 3d$_{5/2}$,
4s$_{1/2}$, 2g$_{7/2}$, 3d$_{3/2}$ and, only to some
extent, for 1j$_{15/2}$. The position 
of the 2g$_{9/2}$, 1i$_{11/2}$ and 1j$_{15/2}$ states is in good agreement
with experiment, while the 3d$_{5/2}$, 1j$_{15/2}$ and 2g$_{7/2}$ orbitals lie
too high in energy. 

The present
calculation is similar to the one of Ref. \cite{colo10} but
the energy shifts are larger, as in the case of $^{40}$Ca,
due probably to the contribution
of non-collective states. We can also make an overall comparison 
with the various results reported in Ref. \cite{CM}, by evaluating an 
average particle-hole gap defined as
\begin{equation}
\Delta\omega = \langle \epsilon_p \rangle - \langle
\epsilon_h \rangle,
\end{equation}
with
\begin{eqnarray}
\langle \epsilon_p \rangle & = & \frac{\sum_{\rm unocc}
(2j+1)\epsilon_{nlj}}{\sum_{\rm unocc}(2j+1)}, \nonumber \\
\langle \epsilon_h \rangle & = & \frac{\sum_{\rm occ}
(2j+1)\epsilon_{nlj}}{\sum_{\rm occ}(2j+1)},
\end{eqnarray}
where the labels ``unocc'' and ``occ'' refer
to the unoccupied and occupied valence shell, respectively.
Starting from the HF value, 9.34 MeV, the HF-PVC value is reduced to
7.89 MeV and gets closer to the experimental value of 6.52 MeV; the difference
between the two values, namely -1.45 MeV, compares well
with the values presented in Ref. \cite{CM}, that range
between -3.2 MeV and -1.1 MeV.

The processes leading to the fragmentation of the single-particle
strength for the orbitals lying far from the Fermi energy  in $^{208}$Pb have
been already extensively discussed within the framework of more
phenomenological studies \cite{CM}; however, for the 
convenience of the reader, in the following we present some
details of the present calculation. 

In many cases, the strong broadening of the single-particle strength observed
in Fig. \ref{lvdPb208} is caused mostly by the coupling with the ISGQR and ISGOR,
due to the favourable matching with the difference between the relevant single-particle states. 
This is the case for the orbitals $(1\rm s_{1/2})^{-1}$ and 
$(2\rm s_{1/2})^{-1}$, which couple to  
$\rm d_{3/2},\ \rm d_{5/2}\otimes 2^+$ 
and to $\rm f_{5/2},\ \rm f_{7/2}\otimes 3^-$. 
The contribution of the $4^+$ strength is small, but not
completely negligible. The main configurations contributing to
the large broadening of the $(3\rm 
s_{1/2})^{-1}$are $\rm f_{5/2},\ \rm f_{7/2}\otimes 3^-$.
In the case of $(1\rm p_{1/2})^{-1}$ and $(2\rm p_{1/2})^{-1}$, 
the $2^+$, $3^-$ and $4^+$ phonons give comparable contributions 
to the strength fragmentation, which is also quite large.
In the case of $(3\rm p_{1/2})$ and $(3\rm p_{3/2})^{-1}$, there
is no good match  with the energy of available single-particle configurations,
and this explains the small amount of fragmentation that
characterizes these states. 
 
In the case of $(1\rm d_{3/2})^{-1}$, the $2^+$, $3^-$ and 
$4^+$ phonons give comparable contributions 
to the fragmentation. In the case of $(2\rm d_{3/2})^{-1}$, 
instead, the $3^{-}$ phonons play the most important
role for the fragmentation: in fact, the relevant  
single-particle configurations are $(3\rm p_{3/2})^{-1}$, 
$(2\rm f_{5/2})^{-1}$, $(2\rm f_{7/2})^{-1}$ and $(1\rm h_{9/2})^{-1}$ 
coupled with the low-lying $3^-_1$ state.
A similar pattern holds for the spin-orbit partners, that is, in 
the case of $(1\rm d_{5/2})^{-1}$, the $2^+$, $3^-$ and $4^+$ 
phonons give comparable contributions
for the fragmentation but in the case of $(2\rm d_{5/2})^{-1}$, 
the $3^{-}$ phonons play the main role for the fragmentation
(with some contribution arising from coupling with 
$4^+$ phonons). 
For $(2\rm d_{3/2})^{-1}$, the 
single-particle configurations involved are 
$(3\rm p_{3/2})^{-1}$, $(2\rm f_{5/2})^{-1}$, $(2\rm f_{7/2})^{-1}$ 
and $(1\rm h_{9/2})^{-1}$ (coupled with the $3^-_1$ state),
as well as $(1\rm i_{13/2})^{-1}$.
In the case of $(1\rm f_{5/2})^{-1}$ and $(1\rm f_{7/2})^{-1}$, 
the $2^+$, $3^-$ and $4^+$ phonons give similar contributions 
to the strength fragmentation; however, the main configuration 
involved turns out to be 
$(1\rm i_{13/2})^{-1}\otimes 3^-$. 
As already mentioned, the state $(2\rm f_{5/2})^{-1}$ is not 
affected much by the particle-vibration coupling. 
In the case of the states $(1\rm g_{7/2})$ and $(1\rm h_{9/2})^{-1}$, 
the $3^-$ is the main responsible for the couplings; however, 
the fragmentation is rather small, and the energy shift is also small. 
In the case of $(1\rm g_{9/2})^{-1}$, the $2^+$, $3^-$ and $4^+$ 
phonons give comparable effects. 
In the case of $(1\rm h_{11/2})^{-1}$, the fragmentation is caused 
by the coupling with the configurations $(3\rm p_{3/2})^{-1}$, 
$(2\rm f_{5/2})^{-1}$, $(2\rm f_{7/2})^{-1}$
and $(1\rm h_{9/2})^{-1}$ $\otimes$  4$^+$. The state $2\rm h_{11/2}$ 
is a resonant state in the continuum: $2^+$ and $3^-$ give the main 
contributions to fragment its strength:  
$1\rm j_{15/2}\otimes 2^+$, 
$3\rm d_{5/2},\ 2\rm g_{7/2},\ 2\rm g_{9/2},\ 1\rm i_{11/2} 
\otimes 3^-$ are 
the main states that produce the strength fragmentation. 
Also $1\rm j_{13/2}$ is a resonant state in the continuum: 
in this case, $3^-$ and $5^-$ are the most relevant phonons
for the fragmentation of the strength: the main configurations
are $2\rm g_{7/2}$, $2\rm g_{9/2}$, $1\rm i_{11/2}\otimes 3^-$ 
and $3\rm d_{3/2}$, $3\rm d_{5/2}$, $2\rm g_{7/2}$, $2\rm 
g_{9/2}$, $1\rm i_{11/2}\otimes 5^-$. Finally, in the case of 
the state $1\rm j_{15/2}$, the fragmentation is mainly caused 
by the coupling with the configurations $1\rm i_{11/2}, 2\rm g_{9/2}
\otimes 3^-$. 
Once more, from considerations related to the matching of
initial and intermediate energies, 
we expect that the low-lying $3^-$ state is the main
contributor. 

\begin{figure}[htbp]
\includegraphics[width=\textwidth,angle=-90,scale=0.8]{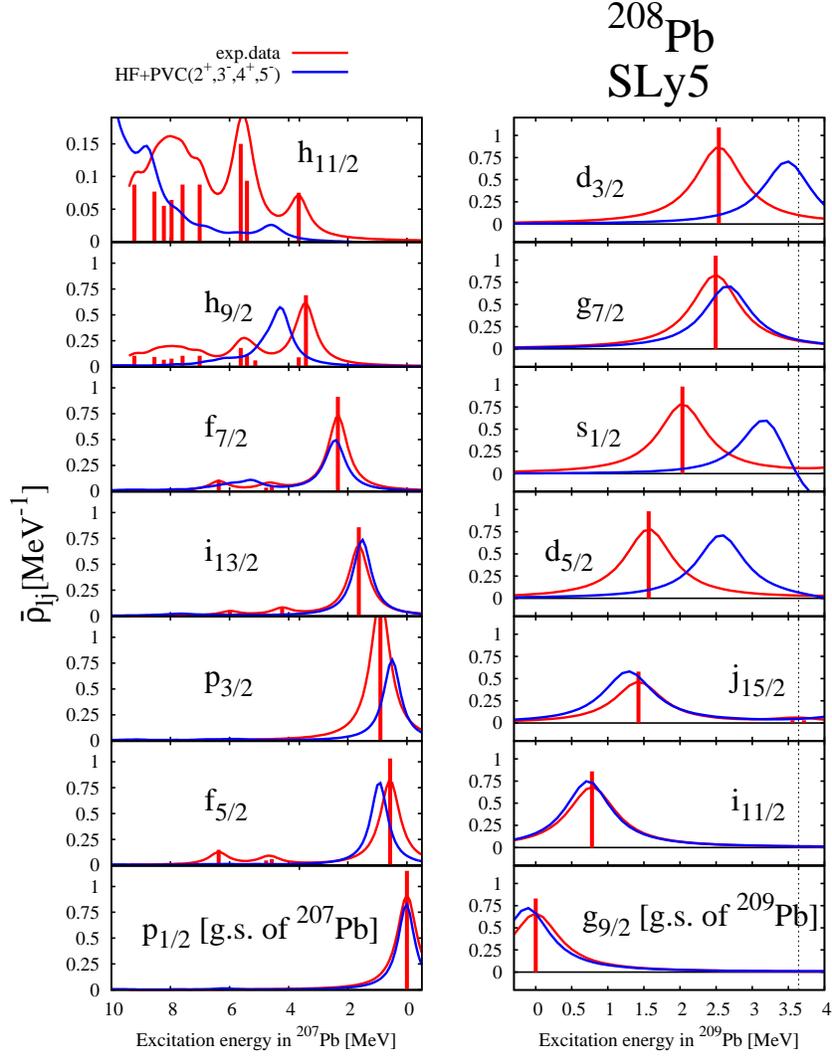}
\caption{(Color online) The same as Fig. \ref{compexp_Ca40} 
in the case of $^{208}$Pb. The experimental data are taken 
from \cite{nucldataA207,nucldataA209}.}
\label{compexp_Pb208}
\end{figure}

\begin{table}[bh!]
\renewcommand\arraystretch{1.5}
\begin{tabular}{ccccccc}
\hline
\multicolumn{7}{c}{$^{208}$Pb} \\
\hline
\multicolumn{3}{c}{Holes} & & \multicolumn{3}{c}{Particles} \\
& \multicolumn{2}{c}{$S_{lj}(^{207}$Pb)} & & &
\multicolumn{2}{c}{$S_{lj}(^{209}$Pb)} \\
\cline{2-3}\cline{6-7}
$J^\pi$ & Exp. & Theory & & $J^\pi$ & Exp. & Theory \\
p$_{1/2}$  & 1.07  & 0.82 & & g$_{9/2}$ & 0.76 & 0.77  \\
p$_{3/2}$  & 1.50  & 0.84 & & s$_{1/2}$ & 0.87 &  0.47 \\
f$_{5/2}$  & 1.07  & 0.84 & & d$_{3/2}$ & 0.93 &  0.52 \\
f$_{7/2}$  & 1.02  & 0.84 & & d$_{5/2}$ & 0.85 &  0.75\\
h$_{9/2}$  & 1.53  & 0.86 & & g$_{7/2}$ & 0.90 &  0.74\\
h$_{11/2}$ & 0.69  & 0.39 & & i$_{11/2}$ & 0.70 & 0.82 \\
i$_{13/2}$ & 0.90  & 0.87 & & j$_{15/2}$ & 0.54 &  0.71\\
\hline
\end{tabular}
\caption{The same as Table \ref{num_compexp_Ca40_sum} for $^{208}$Pb.}
\label{spectr_pb208} 
\end{table}

The cumulative level densities of the various orbitals are compared in Fig. \ref{compexp_Pb208} 
to spectroscopic factors obtained from the  
$^{208}\mbox{\rm Pb}(^{3}\mbox{\rm He},\alpha)$ stripping reaction for hole states,
and from $^{208}\mbox{\rm Pb}(\rm d,\rm p)$ reaction 
\cite{nucldataA207,nucldataA209} for particle states.
The integrated level density is compared to the sum of experimental 
spectroscopic factors in Table \ref{spectr_pb208}. 
The quasiparticle character of the orbitals lying close
to the Fermi energy is a general result of our adopted 
theoretical framework. Our results are in fair overall 
agreement with the experimental findings from transfer 
reactions, which were able to locate most of the 
quasiparticle strength associated with the orbitals lying close 
to the Fermi energy. The quality of the agreement 
varies from one case to the other, and it is hard
to decide whether this has to be attributed either to specific
features of our model, or to deficiencies in the
experimental extraction of the spectroscopic factors
(as testified, e.g., by the fact that some of them
exceed the maximum allowed value of one). Furthermore, 
we must recall that 
(e,e$^\prime$p) experiments lead to much smaller spectroscopic 
factors, and that the relationship between the two kinds of
experiments, as well as the reletive role 
of long- and short-range correlations, is a matter which continues to be 
actively debated \cite{kramer,dickhoff}.   
Last but not least, the very possibility of extracting a
spectroscopic factor as a true observable, has been
recently questioned \cite{hammer,jennings}.

\begin{figure}[htbp]
\includegraphics[width=\textwidth,angle=-90,scale=0.9]{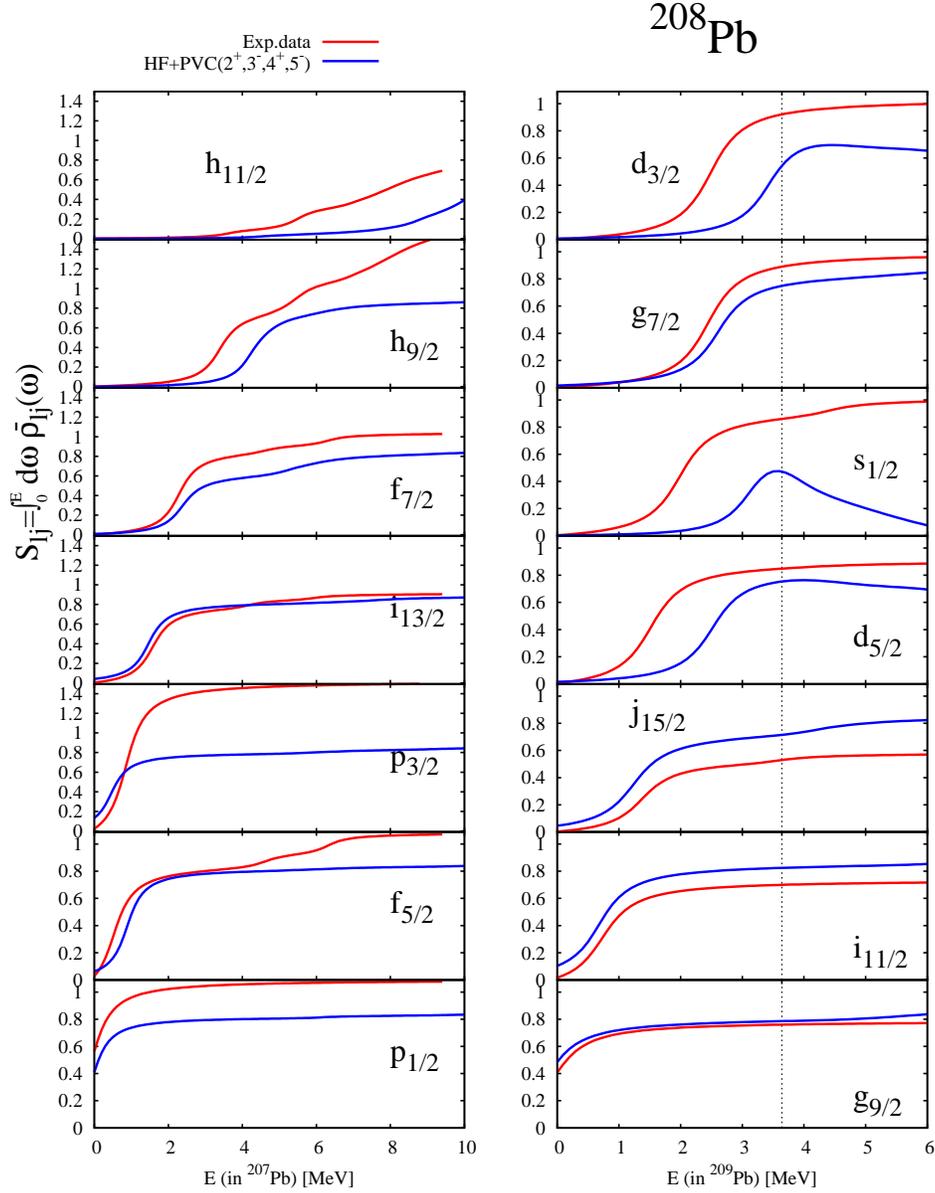}
\caption{(Color online) The same as Fig. 
\ref{compexp_Ca40_sum} in the case of $^{208}$Pb.}
\label{compexp_Pb208_sum}
\end{figure}


\subsection
{Results for $^{24}$O}

In this subsection, we finally give results for $^{24}$O, as  
an example of neutron-rich, weakly bound ($S_n = $ 4.1 MeV) nucleus. 
This nucleus is a doubly magic 
isotope, due to to the usual proton shell closure at Z=8 and
to an ``exotic'' neutron shell gap appearing at N=16.
The magicity of
$^{24}$O, had been suggested by theoretical studies \cite{magic24a,magic24b,magic24c},
and has been established by the measurement of the
(unbound) $^{25}$O ground-state and of its decay spectrum to $^{24}$O, and by 
the extraction of the N=16 single-particle gap from
the $^{23,24,25}$O ground-state energies \cite{Hof08,Hof09}. 
As already mentioned, having a tool that allows studying 
weakly-bound systems by taking proper care of the continuum,
is one of the main motivations of the present work. Interestingly
enough, it has been suggested that nuclei in which the neutron separation energy 
becomes smaller than the proton separation energy 
are characterized by larger single-particle spectroscopic
factors or, in other words, by more pure single-particle
states. This is the feature emerging by the plot shown
in Fig. 6 of Ref. \cite{Gade} and consequently, one of those
that can be analyzed within our framework. We will come back to this
point below.

\begin{figure}[htbp]
\includegraphics[width=\textwidth,angle=-90,scale=0.7]{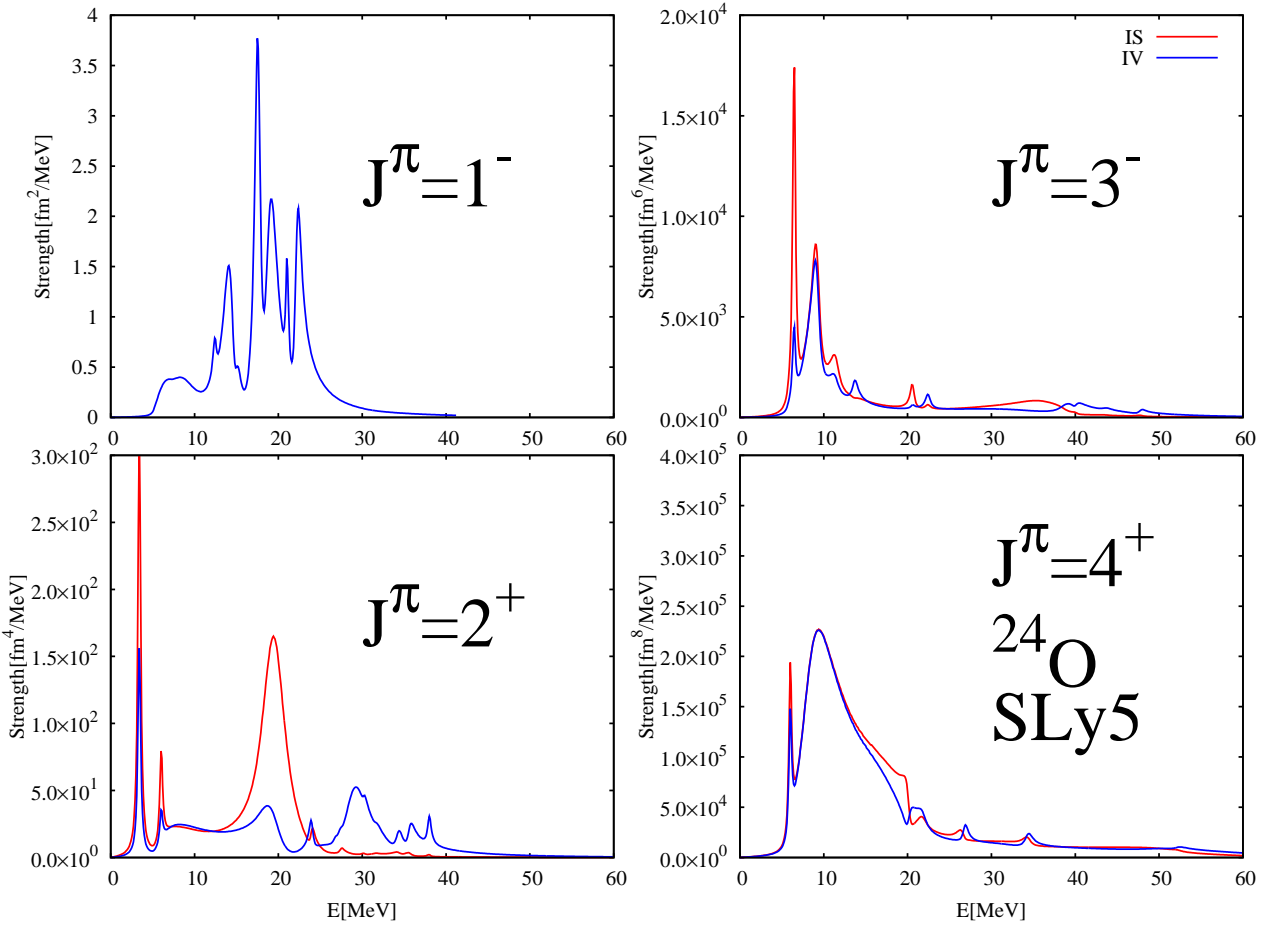}
\caption{(Color online) The RPA strength functions in $^{24}$O 
obtained with the present continuum RPA.}
\label{str_O24}
\end{figure}

In Table \ref{splvtbO24} we provide the HF single-particle
spectrum for neutrons. In Fig. \ref{str_O24}, we illustrate
our results for the RPA strength functions.
Experimental information, although scarce, is available.
The main results are that (i) there should be no bound
excited state \cite{Stan04}, and (ii) the lowest excitation 
should be a 2$^+$ state lying at 4.72 MeV \cite{Hof09}. In
our RPA spectra, the lowest peak among those found 
for the chosen multipolarities is indeed a 2$^+$ one,
and its energy and electromagnetic 
transition probability are 3.4 MeV and
4.2 e$^2$ fm$^4$. 
Experiment has also provided indications for 
the existence of a $1^+$ state lying at 5.33 MeV, but 
our calculations are limited to natural-parity states.  
In the case of 
this nucleus we compute the 1$^-$ strength as well. To ensure that
coupling with 1$^-$ phonons does not introduce any
error associated with spurious strength associated
with the translational mode, we follow the procedure
already discussed in our previous paper \cite{spurious}.
We find a significant amount of dipole strength
lying at energies somewhat below the usual (IS and IV)
giant dipole resonances.


\begin{table}[htbp]
\renewcommand\arraystretch{1.5}
\begin{tabular}{ccccc}
\hline
\hline
Nucleus &\multicolumn{2}{c}{hole states[MeV]}\hspace{0.1cm} 
& \multicolumn{2}{c}{particle states[MeV]} \\[0.5mm]
\hline
$^{24}$O
&1s$\frac{1}{2}$&-39.2&  1d$\frac{3}{2}$ & -1.3 \\
&2s$\frac{1}{2}$& -5.3& & \\
&1p$\frac{1}{2}$&-17.2& & \\
&1p$\frac{3}{2}$&-22.0& & \\
&1d$\frac{5}{2}$& -7.5& & \\
\hline
\end{tabular}
\caption{Skyrme Hartree-Fock single-particle 
energies for $^{24}$O with the interaction SLy5. Hole and 
particle states at negative energies are displayed.}
\label{splvtbO24}
\end{table}

\begin{figure}[htbp]
\includegraphics[width=\textwidth,angle=-90,scale=0.7]{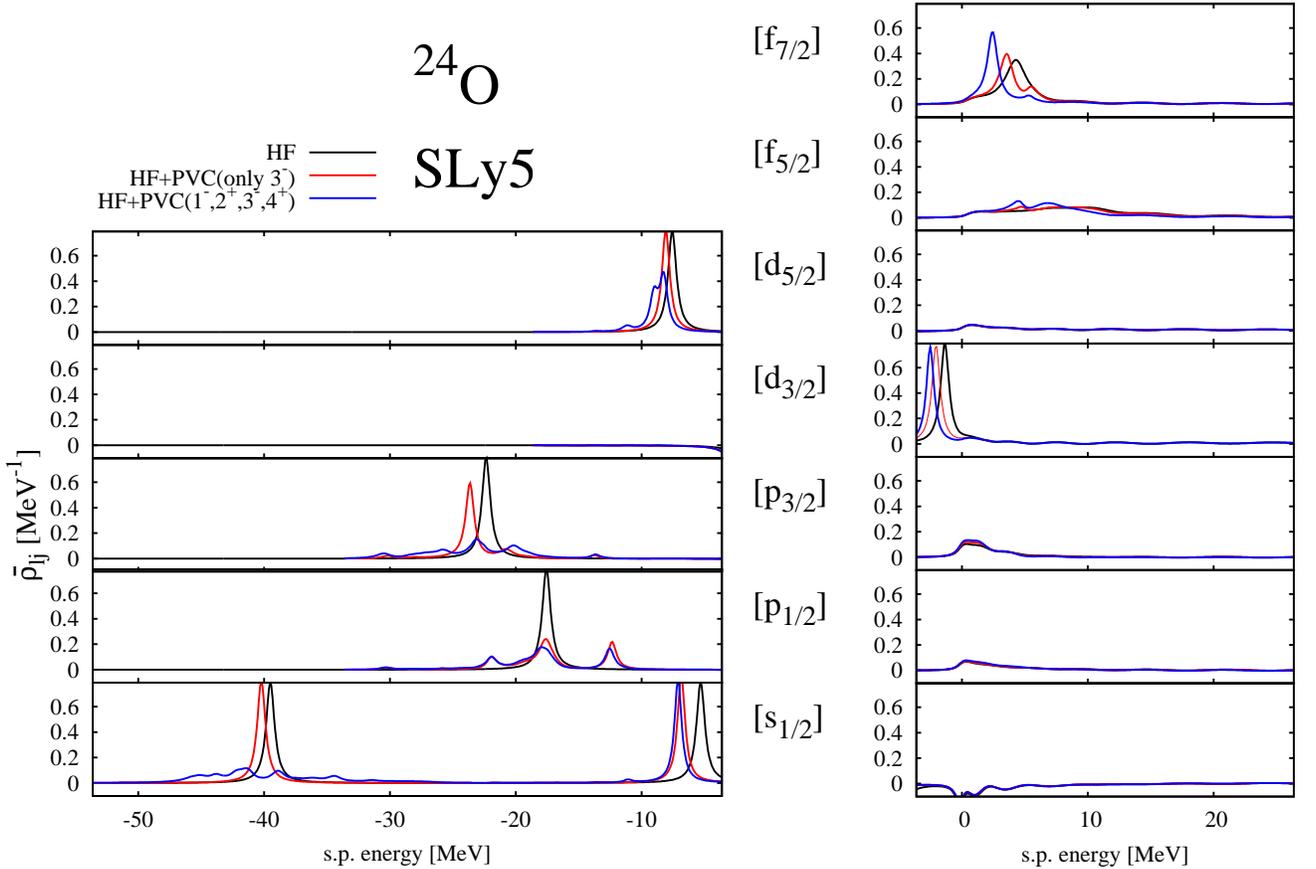}
\caption{(Color online) The same as Fig. \ref{lvdCa40} in the case of 
$^{24}$O. In this case not only 2$^+$, 3$^-$ and 4$^+$ phonons but
1$^-$ phonons as well are considered in the case of the full PVC
result (blue curve).
}
\label{lvdO24}
\end{figure}

In Fig. \ref{lvdO24}, we display our results for the level 
density of $^{24}$O. Before entering into some detail,
we discuss the main emerging features and compare with
what is known experimentally.  
In our 
calculation, the 1d$_{3/2}$ and 2s$_{1/2}$ states have
a marked quasiparticle character, namely they are
associated with a single narrow peak.
It makes sense, therefore, to compare the experimental
value of the gap with the HF and HF-PVC results 
for the energy difference between the 1d$_{3/2}$ and 2s$_{1/2}$
states, that is, 4.0 and 4.6 MeV respectively. The HF-PVC result
is in good agreement with the experimental value of 4.86 MeV. 
In heavy nuclei,
as a rule, the PVC shrinks the single-particle gap
and increases the effective mass (cf. the previous
subsection), but this is not the case in light nuclei
due to the specific effect of having only low 
angular momentum occupied states (as already noticed
in Ref. \cite{colo10}). 
In the present case, while the PVC
pushes the 1d$_{3/2}$ orbital closer to the Fermi energy
(-2.5 MeV compared to the HF value of -1.3 MeV), 
the 2s$_{1/2}$ hole state
is pushed further from it (-7.1 MeV compared to the HF 
value of -5.3 MeV). 

The peak energies of the other orbitals
obtained by using HF-PVC (HF) read 2.4 MeV (4.3 MeV) 
for 1f$_{7/2}$, 
and -8.3 MeV (-7.5 MeV) for 1d$_{5/2}$. The net effect of PVC 
is a shift down of the states. The absolute value of
the energies is expected to depend on the choice of
the effective force. Skyrme forces, as other mean-field
frameworks, tend to predict larger binding in light
neutron-rich nuclei as compared with the experimental
findings, as it is clearly testified by the fact that $^{28}$O 
turns out to be bound in many of these models. In the
present case, the 1d$_{3/2}$ state is bound while it
should be a resonant state. We can nonetheless look
at relative energy differences. The known states 
in $^{23}$O taken from Ref. \cite{Shi07} are, in
addition to the 1/2$^+$ ground-state, a 5/2$^+$ 
state at 2.79 MeV and a 3/2$^+$ at 4.04 MeV (leaving
aside the state at 5.34 whose character is not 
clear, being either 3/2$^-$ or 7/2$^-$). 
These are states that can decay to the $^{22}$O ground
state. In our calculation we can identify states below
the energy threshold for this kind of decay: in particular 
the first 5/2$^+$ state lies at 1.2 MeV in our 
calculation. 

We now discuss, the couplings
that produce fragmentation of most of the single-particle
strength distributions. At variance with the state 2s$_{1/2}$, the state
1s$_{1/2}$ is strongly fragmented. 
This fragmentation 
is chiefly caused by the configurations 1d$_{5/2}^{-1}\otimes\ 2^+$ 
and $( 1\rm p_{1/2})^{-1},\
(1\rm p_{3/2})^{-1}\otimes 1^-$. 
By inspecting the energy difference, we can assume that the IVGQR 
and IVGDR play the main 
role for the fragmentation. 
The state $(1\rm p_{1/2})^{-1}$ is fragmented due to the 
coupling with the configuration $(1\rm d_{5/2})^{-1}\otimes 3^-$: we 
expect that the ISGOR and IVGOR 
play the main role by considering the energy matching. 
$(2\rm s_{1/2})^{-1}\otimes 1^-$ can also contribute to the 
fragmentation, yet to a minor extent. The main configuration 
giving rise to the fragmentation of the state 
$(1\rm p_{3/2})^{-1}$ is the configuration 
$(1\rm d_{5/2})^{-1}\otimes 1^-$. Here, due to the energy 
difference between the hole states $(1\rm p_{3/2})^{-1}$ 
and $(1\rm d_{5/2})^{-1}$, the dipole excitations around $14$ 
MeV do play the main role. 
In the case of the fragmentation 
of the $(1\rm d_{5/2})^{-1}$ state, the configuration 
$(2\rm s_{1/2})^{-1}\otimes 2^+$ is the most important one, and 
the low-lying $2^+$ state at 3.4 MeV is the most relevant. 
Finally, 
the energy shift of the state $1\rm f_{7/2}$ is mostly 
caused by the coupling with the configuration $1\rm d_{3/2} \otimes 3^-$.



\begin{table}
\renewcommand\arraystretch{1.5}
\begin{tabular}{ccccccc}
\hline
\multicolumn{7}{c}{$^{24}$O} \\
\hline
\multicolumn{3}{c}{Holes} & & \multicolumn{3}{c}{Particles} \\
& \multicolumn{2}{c}{$S_{lj}(^{24}$O)} & & &
\multicolumn{2}{c}{$S_{lj}(^{24}$O)} \\
\cline{2-3}\cline{6-7}
$J^\pi$ & Exp. & Theory & & $J^\pi$ & Exp. & Theory \\
s$_{1/2}$  & ---  & 0.81 & & d$_{3/2}$ & --- & 0.83  \\
d$_{5/2}$  & ---  & 0.78 & &  &  &   \\
\hline
\end{tabular}
\caption{The same as Table \ref{num_compexp_Ca40_sum} for $^{24}$O.}
\label{table_spect_o24}
\end{table}

Last but not least, we come back to the point raised above,
namely that the (quasi-particle-like) states around the Fermi
energy, 1d$_{3/2}$ and 2s$_{1/2}$, are quite pure
(see also Table \ref{table_spect_o24}). In our
calculation, there are not optimal energy matching of those
states with other configurations due to the scarcity of low-lying
collective excitations and large gaps between single-particle state.
More generally, in our calculations the coupling of neutron
states is mainly with the proton component of the phonon 
states (due to the dominance of the neutron-proton interaction).
Therefore, neutron states on neutron-rich nuclei are expected
to be more pure because proton excitations are pushed at
higher energy as the neutron excess increases. 
Recently,
other calculations of the spectroscopic factors in light nuclei
within the coupled-cluster approach have become available 
in Ref. \cite{jensen} (see also the critical discussion
in Ref. \cite{hagen}).    

\section{Summary}\label{summary}

The idea that single-particle strength is 
not systematically pure in atomic nuclei, and that
coupling with other degrees of freedom is quite
relevant, is an old idea in nuclear physics. 
Phenomenological calculations based on 
particle-vibration coupling (PVC) for spherical nuclei
have been performed for several deacades, and they
have been quite instrumental to point out some
of the limitations of the pure mean-field
approach (like, e.g., the enhancement of the
effective mass around the Fermi energy).
Microscopic PVC calculations based on the
consistent use of an effective Hamiltonian, have
become available only recently.

None of the mentioned calculations, to our
knowledge, takes proper care of the continuum.
In our work, for the first time, we have implemented
a scheme based on coordinate-space representation 
in which the single-particle states, the vibrations, and
their coupling are calculated with proper
inclusion of the continuum. This is of special
interest if weakly-bound nuclei close to the
drip lines are to be studied. However, also in
well-bound nuclei the present approach present
advantages in the sense that resonant states can
be properly studied. Transfer to the continuum has
been the subject of several experimental studies.

In stable nuclei we obtain results that are in
overall agreement with previous studies. We can,
at the same time, better describe the fragmentation
of the single-particle strength. The shifts of the
single-particle states around the Fermi energy, with
respect to the HF values, are relatively large (in
keeping also with the fact that we cannot restrict
our coupling to collective states only). We obtain
an overall agreement with experiment in $^{208}$Pb,
while in the case of $^{40}$Ca our results point to
the need of re-fitting the Skyrme interaction that
has been devised to work at the mean-field level and
not beyond it.

We have also applied our model to a neutron-rich
nucleus, namely $^{24}$O. This is a double-magic nucleus,
and there are few low-lying states. Because of this,
and also since the neutron states
energy would be more affected by coupling with protons,
the neutron single-particle strength around the Fermi
energy is quite pure (i.e., spectroscopic factors are
rather close to one). While this is in agreement with some
experimental findings, certainly more detailed spectroscopic
studies are needed to extract a global trend and a 
firm understanding.

\begin{acknowledgments}
We thank F. Barranco for discussions that initiated the present work, and
P.F. Bortignon for useful suggestions. 
This work has been partly supported by the Italian Research Project 
"Many-body theory of nuclear systems and implications on the physics 
of neutron stars" (PRIN 2008)".
\end{acknowledgments}

\appendix

\section{Hartree-Fock Green's function}
\label{crgreen}

As it is stated in the main text, it is necessary to use the {\it causal}
Green's function for the Dyson equation because this equation is based on 
the use of the Wick's theorem, and the Wick's theorem applies only to 
time-ordered products. On the other hand, the continuum HF Green's function 
is given in the form of a {\it retarded} function. So we need to compute the 
{\it causal} Green's function starting from the {\it retarded} Green's 
function in order to use the continuum HF Green's function in the Dyson equation. 

The causal HF Green's function is defined by
\begin{eqnarray}
iG_0^C(\bvec{r}\sigma t,\bvec{r}'\sigma' t')
&\equiv&
\bra\Psi_0|\mbox{T}\{
\hat{\psi}(\bvec{r}\sigma t)
\hat{\psi}^\dagger(\bvec{r}'\sigma't')
\}|\Psi_0\ket
\\
&=&
\theta(t-t')
\bra\Psi_0|
\hat{\psi}(\bvec{r}\sigma t)
\hat{\psi}^\dagger(\bvec{r}'\sigma't')
|\Psi_0\ket
-
\theta(t'-t)
\bra\Psi_0|
\hat{\psi}^\dagger(\bvec{r}'\sigma't')
\hat{\psi}(\bvec{r}\sigma t)
|\Psi_0\ket.
\nonumber\\
\end{eqnarray}
The retarded Green's function is instead defined by
\begin{eqnarray}
iG^R_0(\bvec{r}\sigma t,\bvec{r}'\sigma' t')
&\equiv&
\theta(t-t')
\bra\Psi_0|
\{
\hat{\psi}(\bvec{r}\sigma t),
\hat{\psi}^\dagger(\bvec{r}'\sigma't')
\}|\Psi_0\ket.
\end{eqnarray}
From these two definitions, and using $\theta(t'-t)=1-\theta(t-t')$, 
we can find that
\begin{eqnarray}
iG^C_0(\bvec{r}\sigma t,\bvec{r}'\sigma' t')
=
iG^R_0(\bvec{r}\sigma t,\bvec{r}'\sigma' t')
-
\sum_h e^{-\frac{i}{\hbar}e_h(t-t')}
\phi_h^*(\bvec{r}'\sigma')
\phi_h(\bvec{r}\sigma)
\label{RandN}
\end{eqnarray}

The Fourier transform of Eq. (\ref{RandN}) is  
expressed as
\begin{eqnarray}
G^C_0(\bvec{r}\sigma,\bvec{r}'\sigma';\omega)
&=&
G^R_0(\bvec{r}\sigma,\bvec{r}'\sigma';\omega)
+
\sum_h 
\frac{2 i\eta}{(\omega-e_h)^2+\eta^2}
\phi_h(\bvec{r}\sigma)
\phi_h^*(\bvec{r}'\sigma')
\\
&&\mbox{(in the limit $\eta\to 0$)}
\\
&\to&
G^R_0(\bvec{r}\sigma,\bvec{r}'\sigma';\omega)
+
2\pi i
\sum_h \delta(\omega-e_h) 
\phi_h(\bvec{r}\sigma)
\phi_h^*(\bvec{r}'\sigma')
\label{RandN2}
\end{eqnarray}

The continuum HF Green's function $G_{0,lj}$ given by Eq. (\ref{cGF}) is 
regular for the complex energy $E$. So the retarded Green's 
function with a smearing width $\eta$ is expressed as
\begin{eqnarray}
G^R_{0,lj}(rr',\omega)=G_{0,lj}(rr';\omega+i\eta),
\end{eqnarray}
where $\omega$ is the real part of the complex energy. Then the continuum causal 
Green's function can be expressed by
\begin{eqnarray}
G^C_{0,lj}(rr',\omega)=G_{0,lj}(rr';\omega+i\eta)
+
\sum_{n_hl_hj_h} 
\frac{2 i\eta}{(\omega-e_{n_hl_hj_h})^2+\eta^2}
\phi_{n_hl_hj_h}(r)
\phi_{n_hl_hj_h}(r').
\end{eqnarray}

\section{Unperturbed response function}
\label{unpressp}

In general, the RPA theory can be formulated in two ways. One is based on 
the {\it causal} function, while another one on the {\it retarded} function. 
Both formulations give the same results for the physical part of the
spectrum, namely for positive excitation energy. However, a complete 
RPA basis must include negative energy states and the two aforementioned
formulations are different for negative energies.
 
In order to construct the self-energy function of Eq. (\ref{selfen}), we need 
to consider the RPA response function not only in the positive energy domain but 
also in the negative energy domain due to the required energy integration. 
So we cannot use the {\it retarded} RPA response function for the self-energy function. 
The RPA equation for the response function is given by $R=R_0+R_0\kappa R$
in any of the representations, $R_0$ being the unperturbed response function, and $R$ 
the RPA response function. In order to obtain the {\it causal} RPA response function 
by solving this equation, the {\it causal} unperturbed response function should 
be used. 

Normally the continuum (Q)RPA is formulated by using {\it retarded} functions within the  
linear response theory. The continuum HF(B) Green's function is used to build the   
(retarded) unperturbed response function \cite{sagawa,matsuo1}. 
It is therefore necessary to know how to convert the retarded unperturbed response function 
to the causal function in the continuum RPA formalism. We show it in the present
Appendix.

The {\it causal} and {\it retarded} response function in RPA are defined by
\begin{eqnarray}
iR_{0}^C(\bvec{r}t,\bvec{r}'t')
&=&\bra \Psi_0|\mbox{T}
\{\delta\hat{\rho}(\bvec{r}t)\delta\hat{\rho}(\bvec{r}'t')\}
|\Psi_0\ket
\\
&=&
\theta(t-t')
\bra \Psi_0|
\delta\hat{\rho}(\bvec{r}t)\delta\hat{\rho}(\bvec{r}'t')
|\Psi_0\ket
+\theta(t'-t)
\bra \Psi_0|
\delta\hat{\rho}(\bvec{r}'t')
\delta\hat{\rho}(\bvec{r}t)
|\Psi_0\ket
\\
iR_{0}^R(\bvec{r}t,\bvec{r}'t')
&=&
\theta(t-t')
\bra \Psi_0|
[\delta\hat{\rho}(\bvec{r}t),
\delta\hat{\rho}(\bvec{r}'t')]
|\Psi_0\ket,
\end{eqnarray}
respectively (here $|\Psi_0\ket$ is the HF ground state). From these definitions, 
one can find the relation between them as follows,
\begin{eqnarray}
iR_{0}^C(\bvec{r}t,\bvec{r}'t')
&=&
iR_{0}^R(\bvec{r}t,\bvec{r}'t')
+
\bra \Psi_0|
\delta\hat{\rho}(\bvec{r}t)
\delta\hat{\rho}(\bvec{r}'t')
|\Psi_0\ket
\\
&=&
iR_{0}^R(\bvec{r}t,\bvec{r}'t')
-
\sum_{hh'}e^{i(e_h-e_{h'})(t-t')}
\phi_h^*(\bvec{r})\phi_{h'}(\bvec{r})
\phi_{h'}^*(\bvec{r}')\phi_{h}(\bvec{r}').
\end{eqnarray}

The Fourier transformation of the latter equation gives
\begin{eqnarray}
R_{0}^C(\bvec{r}\bvec{r}';\omega)
&=&
R_{0}^R(\bvec{r}\bvec{r}';\omega)
-
\sum_{hh'}
\frac{2 i\eta}{(\omega-e_{h'}+e_{h})^2+\eta^2}
\phi_h^*(\bvec{r})\phi_{h'}(\bvec{r})
\phi_{h'}^*(\bvec{r}')\phi_{h}(\bvec{r}')
w\\
&&(\mbox{in the limit of }\eta\to 0)
\\
&\to&
R_{0}^R(\bvec{r}\bvec{r}';\omega)
-
2 \pi i
\sum_{hh'}
\delta(\omega-e_{h'}+e_{h})
\phi_h^*(\bvec{r})\phi_{h'}(\bvec{r})
\phi_{h'}^*(\bvec{r}')\phi_{h}(\bvec{r}'),
\end{eqnarray}
where $R_{0}^R(\bvec{r}\bvec{r}';\omega)$ can be expressed by means of the retarded 
HF Green's function as
\begin{eqnarray}
R_{0}^R(\bvec{r}\bvec{r}';\omega)
&=&
\sum_h
\phi_h^*(\bvec{r})
G_0^R(\bvec{r}\bvec{r}';\omega+e_h)
\phi_h(\bvec{r}')
+
\phi_h^*(\bvec{r}')
G_0^{R*}(\bvec{r}\bvec{r}';-\omega+e_h)
\phi_h(\bvec{r}).
\label{cunpres}
\end{eqnarray}
In the continuum RPA formalism, the continuum HF Green's function is used as $G^R_0$ 
in Eq. (\ref{cunpres}).

\section{Spectral representation of the Green's function and the response function}

Here we show the spectral representations of the HF Green's function and the RPA 
response function (both causal and retarded). The difference will appear in 
the sign of the imaginary part $\eta$. Actually this is very important 
to obtain the proper self-energy function by using the contour integration, because 
this sign differencee produce changes in the position of the poles of the Green's function 
and of the response function on the complex energy plane (this fact is connected with the
fact that the Wick's theorem can be applied only for the causal function, as already
mentioned).

\begin{eqnarray}
G^C_0(\bvec{r}\sigma,\bvec{r}'\sigma';\omega)
&&=
\sum_h
\frac{\phi_h(\bvec{r}\sigma)\phi^*_h(\bvec{r}'\sigma')}{\omega-e_h-i\eta}
+
\sum_p
\frac{\phi_p(\bvec{r}\sigma)\phi^*_p(\bvec{r}'\sigma')}{\omega-e_p+i\eta}
\\
G^{R}_0(\bvec{r}\sigma,\bvec{r}'\sigma';\omega)
&&=
\sum_h
\frac{\phi_h(\bvec{r}\sigma)\phi^*_h(\bvec{r}'\sigma')}{\omega-e_h+i\eta}
+
\sum_p
\frac{\phi_p(\bvec{r}\sigma)\phi^*_p(\bvec{r}'\sigma')}{\omega-e_p+i\eta}
\end{eqnarray}

\begin{eqnarray}
R_C(\bvec{r}\bvec{r}';\omega)
&&=
\sum_\nu
\frac{\bra 0|\hat{\rho}(\bvec{r})|\nu\ket\bra\nu|\hat{\rho}(\bvec{r}')|0\ket}
{\omega-E_\nu+i\eta}
-
\frac{\bra 0|\hat{\rho}(\bvec{r}')|\nu\ket\bra\nu|\hat{\rho}(\bvec{r})|0\ket}
{\omega+E_\nu-i\eta}
\nonumber\\
\\
R_R(\bvec{r}\bvec{r}';\omega)
&&=
\sum_\nu
\frac{\bra 0|\hat{\rho}(\bvec{r})|\nu\ket\bra\nu|\hat{\rho}(\bvec{r}')|0\ket}
{\omega-E_\nu+i\eta}
-
\frac{\bra 0|\hat{\rho}(\bvec{r}')|\nu\ket\bra\nu|\hat{\rho}(\bvec{r})|0\ket}
{\omega+E_\nu+i\eta}
\end{eqnarray}

\section{Spectral representation of the self-energy function}

The self-energy function in the space-time representation is defined by Eq. (\ref{selfenrt}).
If we insert the HF and the RPA results in this definition, then the 
self-energy function can be expressed as
\begin{eqnarray}
\Sigma(\bvec{r}_1\sigma_1t_1,\bvec{r}_2\sigma_2t_2)
&=&
\kappa(\bvec{r}_1)
G_0(\bvec{r}_1\sigma_1 t_1,\bvec{r}_2\sigma_2t_2)
\kappa(\bvec{r}_2)
iR(\bvec{r}_1t_1\bvec{r}_2t_2)
\\
&=&
\frac{1}{i}
\theta(t_1-t_2)
\sum_{p,\nu}
e^{-i(E_\nu+e_p)(t_1-t_2)}
\kappa(\bvec{r}_1)
\delta\rho_\nu(\bvec{r}_1)
\phi_p(\bvec{r}_1\sigma_1)
\phi_p^*(\bvec{r}_2\sigma_2)
\delta\rho_\nu^*(\bvec{r}_2)
\kappa(\bvec{r}_2)
\nonumber\\
&&
-
\frac{1}{i}
\theta(t_2-t_1)
\sum_{h,\nu}
e^{+i(E_\nu-e_h)(t_1-t_2)}
\kappa(\bvec{r}_1)
\delta\rho_\nu^*(\bvec{r}_1)
\phi_h^*(\bvec{r}_1\sigma_1)
\phi_h(\bvec{r}_2\sigma_2)
\delta\rho_\nu(\bvec{r}_2)
\kappa(\bvec{r}_2)
\label{selfenrt2}
\end{eqnarray}

The Fouriter transform of Eq. (\ref{selfenrt2}) gives
\begin{eqnarray}
\Sigma(\bvec{r}\sigma,\bvec{r}'\sigma';\omega)
&=&
\sum_{h,\nu}
\frac{
\phi_h(\bvec{r}\sigma)\delta\rho^*_\nu(\bvec{r})\kappa(\bvec{r})
\phi_h^*(\bvec{r}'\sigma')\delta\rho_\nu(\bvec{r}')\kappa(\bvec{r}')
}{\omega-e_h+E_\nu-i\eta}
\nonumber\\&&\hspace{1cm}
+
\sum_{p,\nu}
\frac{
\phi_p(\bvec{r}\sigma)\delta\rho_\nu(\bvec{r})\kappa(\bvec{r})
\phi_p^*(\bvec{r}'\sigma')\delta\rho^*_\nu(\bvec{r}')\kappa(\bvec{r}')
}{\omega-e_p-E_\nu+i\eta}
\label{selfenrt3}
\end{eqnarray}

\section{Residual interaction within the Landau-Migdal approximation}

Here we show the explicit expression of the residual interaction within the so-called Landau-Migdal 
approximation. This residual force is used in the self-energy function [Eq. (\ref{selfen2})].
\begin{eqnarray}
\label{LM}
\kappa_{qq'}(r)&=&
\del{h_q}{\rho_{q'}}(r)
=\frac{\delta^2 E}{\delta\rho_q\delta\rho_{q'}}
\hspace{1cm}
\mbox{where $q,p$}
\nonumber\\
&=&
\left\{
\begin{array}{cl}
(q=q') &
\displaystyle{
\frac{t_0}{2}(1-x_0)}
\\&
\displaystyle{
+\frac{t_3}{12}\rho^\gamma\Big[ 
(\gamma+2)(\gamma+1)(1+\frac{x_3}{2})
-
(x_3+\frac{1}{2})
\left(
2+4\gamma\frac{\rho_q}{\rho}  
+\gamma(\gamma-1)\sum_\alpha\left(\frac{\rho_\alpha}{\rho}\right)^2
\right)
\Big] 
}
\\&
\displaystyle{
+\frac{1}{4}\left(t_1(1-x_1)+3t_2(1+x_2)\right)k_F^2}
\\&
\\
(q\neq q') &
\displaystyle{
t_0(1+\frac{x_0}{2})}
\\&
\displaystyle{
+\frac{t_3}{12}\rho^\gamma\Big[ 
(\gamma+2)(\gamma+1)(1+\frac{x_3}{2})
-
(x_3+\frac{1}{2})
\left(
2\gamma
+\gamma(\gamma-1)\sum_\alpha\left(\frac{\rho_\alpha}{\rho}\right)^2
\right)
\Big] 
}
\\&
\displaystyle{
+\frac{1}{2}\left(t_1(1+\frac{x_1}{2})+t_2(1+\frac{x_2}{2})\right)k_F^2}
\end{array}
\right.
\end{eqnarray}

\end{document}